\documentclass[12pt]{article}

\usepackage[totalwidth=460truept,totalheight=622truept]{geometry}
\usepackage{latexsym,graphicx,amsfonts,amssymb,amsmath,xcolor,dsfont}
\usepackage[hypertexnames=false,hidelinks]{hyperref}

\def\theequation{\arabic{section}.\arabic{equation}}
\newcommand{\sect}[1]{\setcounter{equation}{0}\section{#1}}
\renewcommand{\theequation}{\thesection.\arabic{equation}}
\global\arraycolsep=1truept
\renewcommand{\theequation}{\arabic{section}.\arabic{equation}}

\newcommand{\be}{\begin{equation}}
\newcommand{\ee}{\end{equation}}
\newcommand{\bea}{\begin{eqnarray}}
\newcommand{\eea}{\end{eqnarray}}

\newcommand{\bs}{\begin{split}}
\newcommand{\es}{\end{split}}

\usepackage[format=hang,font=small,textfont=it]{caption}

\begin{document}

\null

\begin{center}
{\LARGE \textbf{Power Spectra in Double-Field Inflation }}
\vskip.8truecm
{\LARGE \textbf{Using Renormalization-Group Techniques}}

\vskip1truecm

\textsl{{\large Bohdan Grzadkowski$^{ \, a}$ and Marco Piva$^{\, b}$}}

\vskip .2truecm

{\textit{University of Warsaw, Faculty of Physics, \\ Pasteura 5, 02-093 Warsaw, Poland}}

\vspace{0.2cm}

$^{a}$bohdan.grzadkowski@fuw.edu.pl \ \ $^{b}$mpiva@fuw.edu.pl
\vskip1truecm

\textbf{Abstract}
\end{center}

A perturbative strategy for inflation described by two-inflaton fields is developed using a mathematical analogy with the renormalization-group. Two small quantities, $\alpha$ and $\lambda$, corresponding to standard slow-roll parameters are defined and systematic expansions of all inflationary quantities in terms of powers of $\alpha$ and $\lambda$ are found. No other slow-roll parameters are needed. To illustrate this perturbative method in the multi-field context, we adopt a simple two-inflaton model with quadratic potentials in the parameter range where both fields contribute similarly to the dynamics of inflation. 
The model, even though it is not a viable alternative for phenomenological description of the inflationary period, nicely illustrates subtleties of the perturbative approach. In particular, this method allows us to derive two independent gauge-invariant scalar perturbations that are conserved in the superhorizon limit, overcoming typical problems that emerge in multi-field inflation. Furthermore, it is possible to perform nonperturbative resummations that allow to study the model in a true multi-field regime. We derive tensor and scalar power spectra to the next-to-next-to-leading and next-to-leading orders, respectively, as well as their spectral indices. The hierarchy between the two scalar perturbations allows us to single out the dominant entry of the scalar power-spectrum matrix. Modifications due to the second inflaton occur already at the leading order.
Finally, we explain why the quadratic two-inflaton model is not compatible with the present experimental constraints even though non-trivial corrections to scalar perturbations do emerge.

\vfill\eject

%%%%%%%%%%%%%%%%%%%%%%%%%%%%%%%%%%%%%%%%%%%%%%%%%%%%
\sect{Introduction}

The inflationary paradigm~\cite{Brout:1977ix, Starobinsky:1980te, Guth:1980zm, Linde:1981mu} provides an efficient yet simple solution to the main puzzles of primordial cosmology (e.g. the horizon and the flatness problems) and easily accommodates the present observations concerning the cosmic microwave background (CMB)~\cite{Planck:2018jri}.
Such results are obtained assuming that, during the initial phase of the universe, a scalar field with a certain potential has a nontrivial, time-dependent but spatially homogeneous background. Then for the potential that implies slowly-rolling scalar background one finds exponentially growing solution of the Friedmann equations, known as the inflation (quasi-de Sitter).

Finally, quantum perturbations of the scalar field, together with those of the spacetime metric, accounts for the small temperature fluctuations observed in the CMB.

The simplest class of potentials, i.e. monomials, either do not fit the data or reproduce them poorly, and alternatives should be used. One possibility is to add new free parameters by considering polynomial potentials, which weakens the predictive power of a given model. Another way to better fit the data without introducing new free parameters is to consider more exotic potentials that are not polynomials in the fields (see~\cite{Martin:2013tda} for a complete list). However, this choice spoils the simplicity of the models since such potentials are somehow unusual from a particle-physics point of view, where renormalizability and power counting severely restrict the range of possibilities. The only nonpolynomial potential that is allowed by power counting is the Starobinsky one~\cite{Starobinsky:1980te}. In fact, it can be obtained by adding the $R^2$ term to the Hilbert action, which is the only term with dimension 4 that does not introduce ghosts. For renormalizability it is necessary to include also the cosmological constant, which is irrelevant for primordial cosmology, and the $R_{\mu\nu}R^{\mu\nu}$ term~\cite{Stelle:1976gc}. The latter introduces a massive spin-2 ghost. Unitarity can be recovered by using an alternative quantization prescription for the ghost, turning it into a so-called “purely virtual particle"~\cite{Anselmi:2017yux, Anselmi:2017ygm, Anselmi:2018ibi}. Its presence modifies the power spectra~\cite{Anselmi:2020lpp} introducing small deviations from Starobinsky predictions that could be tested in the near future~\cite{Hazumi:2019lys}.  

Yet another possibility considered in the literature is to introduce additional fields that actively participate to inflation dynamics. In fact, the assumptions of homogeneity and isotropy of cosmological spacetime allows for any scalar degree of freedom to have a nontrivial background. Since several beyond-the-standard-model models assume the existence of new scalars, it is interesting to study their possible participation to the inflationary phase. 

Various multi-field models have been studied in the literature (see for example~\cite{Polarski:1992dq, Starobinsky:1994mh, Garcia-Bellido:1995him, Gordon:2000hv, Lalak:2007vi,Avgoustidis:2011em}. However, most of the results rely either on numerical calculations or on the assumption that the background fields follow a specific trajectory in field space. Furthermore, the additional fields are often chosen to be spectators, effectively reducing the model to single-field inflation~\cite{Gundhi:2018wyz, Kubota:2022pit}, and systematic derivations of analytical predictions for truly multi-inflaton cases are usually missing. 

In this paper we study a double-field model with quadratic potential for both fields, denoted by $\chi$ and $\phi$, and use the formalism developed in~\cite{Anselmi:2020shx, Anselmi:2021rye} to obtain analytical and perturbative power spectra. These technique make use of a mathematical correspondence between the background evolution during the inflationary phase and the renormalization-group flow in quantum field theory. In this analogy the small slow-roll parameters play the role of coupling constants, their equations of motion correspond to equations satisfied by running coupling constants as a function the regularization scale $\mu$, its counterpart being the conformal time $\tau$. Finally, the conservation of correlation functions on superhorizon scales is the analog of the Callan-Symanzik equation. These ingredients define a sort of renormalization-group flow that describes the evolution of the universe during the inflationary era. We stress that this is just a mathematical analogy. In particular, the ``beta functions" do not account for quantum corrections. They rather represent the background dynamics during the slow-roll phase in a perturbative way. More details on this approach can be found in~\cite{Anselmi:2021rye}.

The free parameters of the model are the two masses $m_{\chi}$ and $m_{\phi}$ and we do not assume any relation between them. Other known cases, e.g. single-field or spectator-field cases,  can be recovered form our formulas as limiting cases. Moreover, we address the problem of the nonconservation of scalar perturbations, which is a typical issue in multi-field inflation. In fact, when more than one scalar field participate to inflation we can define a gauge-invariant scalar perturbation for each field. However, in general they are not conserved on superhorizon scales, making it more difficult to confront them with experimental data. Thanks to the perturbative techniques adopted here, we are able to define two independent gauge-invariant perturbations that are conserved on superhorizon scales. Moreover, due to the difference in their magnitude, it is possible to identify a dominant component. These features make their power spectra good candidates for observables to be compared with experiments. Then, in a consistent way, order by order in powers of small quantities, parameters of the model could be constraint. 

Finally, we mention that a similar investigation has been done in~\cite{Anselmi:2021dag}, where the theory studied in~\cite{Anselmi:2020lpp} is supplemented by an additional scalar field with a quadratic potential. However, in that case the presence of the second inflaton only affects the subleading corrections, while in the model studied here even the leading order is modified. This is due to certain nonperturbative resummations that can be identified thanks to a symmetry emerging in the model.
 
The paper is organized as follows. In~\autoref{sec:background} we introduce the renormalization-group analogy in the context of the double-field model and derive all the background quantities as perturbative series in two independent slow-roll parameters. In~\autoref{sec:running} we solve the differential equations for the slow-roll parameters. In~\autoref{sec:perturbations} we discuss the strategy that we adopt to derive  conserved perturbations and their power spectra and work them out to the next-to-next-to-leading (NNL) order for tensors and to the next-to-leading (NL) order for scalars.
Finally, \autoref{sec:conclusions} contains our conclusions.

%%%%%%%%%%%%%%%%%%%%%%%%%%%%%%%%%%%%%%%%%%%%%%%%%%%%%%%%%%%%%%%%%%%%%%%%%%%%%%%%%%%%%%
\sect{Double-field model and background equations in the renormalization-group analogy }
\label{sec:background}

In this section we show the background-field equations of the model and introduce the quantities used in the renormalization-group analogy that we adopt through the paper.

The action is
\be\label{eq:actminimal}
S(g,\phi)=\frac{1}{2}\int\mathrm{d}^4x\sqrt{-g}\left[-\frac{1}{\kappa^2}R+\partial_{\mu}\chi\partial_{\nu}\chi g^{\mu\nu}+\partial_{\mu}\phi\partial_{\nu}\phi g^{\mu\nu}-m_{\chi}^2\chi^2-m_{\phi}^2\phi^2\right],
\ee
where $\kappa^2=8\pi G$ and $G$ is Newton constant. Assuming isotropy and homogeneity we consider the background scalar fields that depend exclusively on time and the line-element squared of the Friedmann-Lema\^{\i}tre-Robertson-Walker form 
\be
\mathrm{d}s^2=g_{\mu\nu}\mathrm{d}x^{\mu}\mathrm{d}x^{\nu}=\mathrm{d}t^2-a(t)^2\mathrm{d}\mathbf{x}^2.
\ee
The background-field equations are 
\be
H^2=\frac{\hat{\kappa}^2}{2}\rho, \qquad \dot{H}=-\frac{3}{4}\hat{\kappa}^2(\rho+p),
\ee
\be
\ddot{\chi}+3H\dot{\chi}+m^2_{\chi}\chi=0, \qquad \ddot{\phi}+3H\dot{\phi}+m^2_{\phi}\phi=0,
\ee
where
\be
H(t)=\frac{\dot{a}}{a}, \qquad \hat{\kappa}\equiv\sqrt{\frac{2}{3}}\kappa,
\ee

\be
\rho=\frac{1}{2}\left(\dot{\chi}^2+\dot{\phi}^2+m^2_{\chi}\chi^2+m^2_{\phi}\phi^2\right), \qquad p=\frac{1}{2}\left(\dot{\chi}^2+\dot{\phi}^2-m^2_{\chi}\chi^2-m^2_{\phi}\phi^2\right).
\ee
Since we want to define a perturbative regime where we can apply the renormalization-group analogy mentioned above, we introduce the following slow-roll parameters~\footnote{In the single inflaton ($\phi$) scenario, $\lambda$ would be related to the standard inflation parameter $\varepsilon\equiv -\dot H/H^2$ by $\varepsilon=3\lambda^2$.}
\be\label{eq:defalphalambda}
\alpha\equiv\frac{\hat{\kappa}\dot{\chi}}{2H}, \qquad \lambda\equiv\frac{\hat{\kappa}\dot{\phi}}{2H}
\ee
and express every other quantity as a perturbative series in $\alpha$ and $\lambda$, possibly up to an overall nonpolynomial function. For this purpose it is convenient to introduce also the following quantities

\be\label{eq:vbeta}
v\equiv-\frac{1}{aH\tau}, \qquad \beta_{\alpha}\equiv\frac{\mathrm{d}\alpha}{\mathrm{d}\ln|\tau|},\qquad \beta_{\lambda}\equiv\frac{\mathrm{d}\lambda}{\mathrm{d}\ln|\tau|},
\ee
where $\tau$ is the conformal time
\be
\tau=-\int_{t}^{\infty} \frac{\mathrm{d}t'}{a(t')}.
\ee
In terms of these quantities the background-field equations of motion (EoM) read
\be\label{eq:background1}
\frac{\hat{\kappa}^2}{4}\left(m^2_{\chi}\chi^2+m^2_{\phi}\phi^2\right)=H^2\left(1-\alpha^2-\lambda^2\right),  \qquad \dot{H}=-3H^2(\alpha^2+\lambda^2)
\ee
\be\label{eq:backgroundalpha}
-v\beta_{\alpha}=-3\alpha(1-\alpha^2-\lambda^2)-\frac{\hat{\kappa}m^2_{\chi}\chi}{2H^2},
\ee
\be\label{eq:backgroundlambda}
-v\beta_{\lambda}=-3\lambda(1-\alpha^2-\lambda^2)-\frac{\hat{\kappa}m^2_{\phi}\phi}{2H^2}.
\ee
For future use, we also define
\be
\varrho\equiv \frac{m_{\chi}^2}{m_{\phi}^2}, \qquad \zeta\equiv \varrho+\frac{1}{\varrho}.
\ee

Hereafter we assume $\alpha,\lambda>0$ and we consider slow-roll inflation defined by the condition $\alpha, \lambda\ll 1$. Our first task is to solve background equations of motion 
\eqref{eq:background1}-\eqref{eq:backgroundlambda} in terms of power series in $\alpha$ and $\lambda$. 

We start by assuming a general expansion for the beta functions 
\be
\beta_{\alpha}(\alpha,\lambda)=\sum_{n,m=0}^{\infty}\lambda^n\alpha^mb_{nm}, \qquad \beta_{\lambda}(\alpha,\lambda)=\sum_{n,m=0}^{\infty}\lambda^{n}\alpha^mc_{nm},
\ee
where $b_{nm}$ and $c_{nm}$ are numerical coefficients to be determined. From the results of~\cite{Anselmi:2021rye} we know that in single-field inflation case with a monomial potential the beta function is of order 3. Here that case is recovered by setting either $\alpha$ or $\lambda$ to zero. Therefore, we require
\be\label{eq:betacond}
\beta_{\alpha}(\alpha,0)=\mathcal{O}(\alpha^3), \qquad \beta_{\lambda}(0,\lambda)=\mathcal{O}(\lambda^3),
\ee
which leads to
\be
b_{00}=b_{01}=b_{02}=c_{00}=c_{10}=c_{20}=0.
\ee 

Since, every background quantity depends on time exclusively through $\alpha(t)$ and $\lambda(t)$, it is useful to express the time derivative in terms of partial derivatives with respect to $\alpha$ and $\lambda$ as follows
\be
\frac{\mathrm{d}}{\mathrm{d}t}=-Hv\left(\beta_{\alpha}\frac{\partial}{\partial\alpha}+\beta_{\lambda}\frac{\partial}{\partial\lambda}\right).
\ee
The first equation we need is obtained by applying one time derivative to the definition of $v$ in~\eqref{eq:vbeta}. This gives
\be\label{eq:eqv}
\beta_{\alpha}\frac{\partial v}{\partial\alpha}+\beta_{\lambda}\frac{\partial v}{\partial\lambda}=1-3(\alpha^2+\lambda^2)-v,
\ee 
which can be solved by assuming that $v$ is a power series
\be
v(\alpha,\lambda)=\sum_{n,m=0}^{\infty}\lambda^n\alpha^mv_{nm},
\ee
where $v_{nm}$ are coefficients. Then from the definitions of $\alpha$ and $\lambda$ we obtain
\be \label{eq:eqchiphi}
\beta_{\alpha}\frac{\partial \chi}{\partial\alpha}+\beta_{\lambda}\frac{\partial \chi}{\partial\lambda}=-\frac{2\alpha}{v\hat{\kappa}}, \qquad \beta_{\alpha}\frac{\partial \phi}{\partial\alpha}+\beta_{\lambda}\frac{\partial \phi}{\partial\lambda}=-\frac{2\lambda}{v\hat{\kappa}},
\ee
while from the second equation in~\eqref{eq:background1} we get
\be\label{eq:eqH}
\beta_{\alpha}\frac{\partial H}{\partial\alpha}+\beta_{\lambda}\frac{\partial H}{\partial\lambda}=3(\alpha^2+\lambda^2)\frac{H}{v}.
\ee
It turns out that the last three equations cannot be solved by assuming a power series for $\chi$, $\phi$ and $H$. Instead, we have to adopt the following ansatz 
\be\label{eq:chiexp}
\chi(\alpha,\lambda)=\frac{1}{\alpha}\sum_{n,m=0}^{\infty}\left(\frac{\lambda}{\alpha}\right)^n\alpha^m\chi_{nm},
\ee
\be\label{eq:phiexp}
\phi(\alpha,\lambda)=\frac{1}{\alpha}\sum_{n,m=1}^{\infty}\left(\frac{\lambda}{\alpha}\right)^n\alpha^{m-1}\phi_{nm},
\ee
\be\label{eq:Hexp}
H(\alpha,\lambda)=\frac{1}{\alpha}\sum_{n,m=0}^{\infty}\left(\frac{\lambda}{\alpha}\right)^n\alpha^mh_{nm},
\ee
where $\chi_{nm}$, $\phi_{nm}$ and $h_{nm}$ are coefficients.

We shall note that all formulas must be invariant under the exchanges 
\be\label{eq:alphalambdaexch}
\alpha\leftrightarrow\lambda, \qquad m_{\chi}\leftrightarrow m_{\phi},
\ee
since the role of the two fields can be switched without affecting the results. However, the expansions~\eqref{eq:chiexp}--\eqref{eq:Hexp} seem to violate such property. This apparent asymmetry between $\alpha$ and $\lambda$ is due to the fact that we choose to expand first in $\lambda$ and then in $\alpha$, which implicitly assumes that $\lambda<\alpha$.
In order to recover the symmetry~\eqref{eq:alphalambdaexch} we should resum the powers of $\lambda/\alpha$. After solving the equations~\eqref{eq:eqH} and~\eqref{eq:eqchiphi} for the coefficients $\chi_{nm}$, $\phi_{nm}$ and $h_{nm}$, we find that indeed the powers of $\lambda/\alpha$ can be resumed into geometric series and the expansions have the form
\be
\chi(\alpha,\lambda)=-\frac{2}{3\hat{\kappa} f_{\chi}}\tilde{\chi}(\alpha,\lambda), \qquad \phi(\alpha,\lambda)=-\frac{2}{3\hat{\kappa} f_{\phi}}\tilde{\phi}(\alpha,\lambda),\qquad  H(\alpha,\lambda)=\frac{m_{\chi}}{3 f_H}\tilde{H}(\alpha,\lambda),
\ee
where 
\be
 f_{\chi}\equiv \frac{f_H^2}{\alpha}, \qquad f_{\phi}\equiv\frac{f_H^2}{\varrho\lambda},\qquad f_H\equiv\sqrt{\alpha^2+\varrho\lambda^2},
\ee
and $\tilde{\chi}$, $\tilde{\phi}$ and $\tilde{H}$
are power series that do not depend on the order in which we expand.
To obtain this, we have noticed that
\be
 \frac{\chi_{n0}}{\chi_{00}}=\frac{\phi_{n1}}{\phi_{11}}=(-\varrho)^n, \qquad \frac{h_{n0}}{h_{00}}=\frac{(2n-1)!!}{(2n)!!}(-\varrho)^n
\ee
up to the order 14 and then verified that 
\be
\chi\simeq -\frac{2}{3 \hat{\kappa}}\frac{\alpha}{\alpha^2+\varrho \lambda^2},\qquad \phi\simeq -\frac{2}{3 \hat{\kappa}}\frac{\varrho\lambda}{\alpha^2+\varrho \lambda^2},\qquad H\simeq \frac{m_{\chi}}{3}\frac{1}{\sqrt{\alpha^2+\varrho \lambda^2}}
\ee
are the leading order solutions.

It is important to highlight that recovering the symmetry~\eqref{eq:alphalambdaexch} allows us to treat the model in a true multi-field regime, where there is no distinction between roles of $\chi$ and $\phi$. 

All together we find

\be\label{eq:vseries}
v(\alpha,\lambda)=1-3 \left(\alpha ^2+\lambda ^2\right)-18\left[\alpha^4+\lambda^4+(4-\zeta)\alpha^2\lambda^2\right]+\mathcal{O}(6),
\ee

\be\label{eq:chitildeseries}
\tilde{\chi}(\alpha,\lambda)=1-\alpha ^2-\lambda ^2 \varrho+\alpha ^2 \lambda ^2 \left(3-\frac{1}{\varrho }\right)+\alpha ^4+\lambda ^4 \left(3 \varrho -\varrho ^2-1\right)+\mathcal{O}(6),
\ee

\be\label{eq:phitildeseries}
\tilde{\phi}(\alpha,\lambda)=1-\lambda ^2-\frac{\alpha ^2}{\varrho }+\alpha ^2 \lambda ^2 (3-\varrho )+\lambda ^4+\alpha ^4\left(\frac{3}{\varrho}-\frac{1}{\varrho^2}-1\right)+\mathcal{O}(6),
\ee

\be\label{eq:Htildeseries}
\tilde{H}(\alpha,\lambda)=1-\frac{1}{2}\left(\alpha^2+\lambda^2\right)+\frac{7}{8}\left(\alpha^4+\lambda^4\right)+\frac{1}{4} \alpha ^2 \lambda ^2 \left(19-6\zeta\right)+\mathcal{O}(6),
\ee

\be\label{eq:betaal}
\beta_{\alpha}(\alpha,\lambda)=-3 \alpha ^3+3 \alpha  \lambda ^2 (\varrho -2)+3\alpha ^3 \lambda
   ^2 \left(\varrho-3-\frac{2}{\varrho } \right)+3 \alpha 
   \lambda ^4 \left(\varrho +\varrho ^2-4\right)-6 \alpha ^5+\mathcal{O}(7),
\ee

\be\label{eq:betalam}
\beta_{\lambda}
(\alpha,\lambda)=-3 \lambda ^3+3\lambda  \alpha ^2 \left(\frac{1}{\varrho
   }-2\right)+3\lambda ^3 \alpha ^2
   \left(\frac{1}{\varrho }-3-2 \varrho\right)+3 \lambda  \alpha ^4\left(\frac{1}{\varrho}+\frac{1}{\varrho^2} -4\right)-6 \lambda ^5+\mathcal{O}(7).
\ee
 We have used the notation $\mathcal{O}(N)\equiv\mathcal{O}(\alpha^n\lambda^m)$ with $n+m=N$ to label the higher orders in $\alpha$ and $\lambda$. Note that if we perform the substitutions~\eqref{eq:alphalambdaexch}
the quantities $v$ and $H$ are invariant, while \be
\chi\leftrightarrow\phi, \qquad \beta_{\alpha}\leftrightarrow\beta_{\lambda}
\ee
as it should be. This fact can be used throughout the paper as a consistency check.
We should also point out that the series obtained in this way, as well as those in the next sections, are asymptotic. More details on their optimal truncation are given in~\autoref{app:asymp}.

It is interesting to compare the two-field case with the single-field inflation where the following parameters are usually adopted\footnote{The parameter $\eta_H$ is usually~\cite{Baumann:2022mni} denoted as $\eta$. We introduce the subscript $H$ to avoid confusion with the rescaled conformal time defined in the next section.}
\be
\varepsilon\equiv -\frac{\dot{H}}{H^2}=3(\alpha^2+\lambda^2), \qquad 
\eta_H\equiv \frac{\dot\varepsilon}{H\varepsilon}=-2v\frac{\alpha\beta_{\alpha}+\lambda\beta_{\lambda}}{\alpha^2+\lambda^2}, \qquad 
w \equiv \frac{p}{\rho}=-1+2(\alpha^2+\lambda^2).
\ee

For $\alpha,\lambda \ll 1$, $\varepsilon$ and $w+1$ are small, as expected, regardless if one considers one or two inflatons. However, the case of $\eta_H$ is different. Indeed, we find
\be
\eta_H=
\frac{6}{\alpha^2+\lambda^2}\left[(\alpha^2+\lambda^2)^2+\alpha^2\lambda^2(2-\zeta)+\mathcal{O}(6)\right]=\frac{2}{\varepsilon}\left[\varepsilon^2+9\alpha^2\lambda^2(2-\zeta)+\mathcal{O}(6)\right],
\label{eq:etaH}
\ee
which, for $\zeta\gg 1$, could be large, in contrast with the single-field case where $\eta_H =2\varepsilon + {\cal O}(\varepsilon^2)$, so it is guaranteed to be small for near-de-Sitter geometry~\footnote{It is also worth seeing that $(2-\zeta)=-(\varrho-1)^2/\varrho$, so that this term vanishes for the $O(2)$-invariant limit, i.e. $\varrho=1$. The same happens in higher orders, since $\varrho=1$ is equivalent to single-field inflation.}.

For completeness, we also define the following parameters
\be
\delta_\chi\equiv -\frac{\ddot{\chi}}{H\dot{\chi}}=3\alpha^2+3\lambda^2+v\frac{\beta_{\alpha}}{\alpha},\qquad 
\delta_\phi\equiv -\frac{\ddot{\phi}}{H\dot{\phi}}=3\alpha^2+3\lambda^2+v\frac{\beta_{\lambda}}{\lambda},
\ee
which are related to $\varepsilon$ and $\eta_H$ by
\be
6(\alpha^2\delta_{\chi}+\lambda^2\delta_{\phi})=\varepsilon(2\varepsilon-\eta_H).
\ee
Again, in the single-field case, e.g. $\lambda=0$, the usual relation is recovered
\be
\eta_H=2(\varepsilon-\delta_{\chi}).
\ee

Finally, we mention that $\alpha$ and $\lambda$ are convenient and small inflation parameters when the background quantities are considered. However, it turns out that for  power spectra and spectral indices
it is better to use the following variables
\be\label{eq:xyvariables}
x\equiv\frac{\alpha^2}{\sqrt{\alpha^2+\varrho^2\lambda^2}}, \qquad y\equiv\frac{\varrho\lambda^2}{\sqrt{\alpha^2+\varrho^2\lambda^2}}.
\ee
In fact, using~\eqref{eq:xyvariables} those quantities can be written as power series up to overall factors, while it is not possible using $\alpha$ and $\lambda$ (see~\autoref{sec:perturbations}). 
Nevertheless, we choose to use $\alpha$ and $\lambda$ until the very end, since it is easier to relate them to the standard parameters.

\subsection{On the generality of the approach}
Here, we are going to show how the renormalization-group analogy can be applied in a more general context. In particular, we consider a double-field model with a
non-trivial field-space metric in order to show that the method developed here could be also applied in that case. 

A typical example is given by the following action
\be
\label{eq:EinsteinFrame}
S_E(g,\phi,\chi)=\int\sqrt{-g}\left[-\frac{1}{2\kappa^2}R+\frac{1}{2}\partial_{\mu}\chi\partial_{\nu}\chi g^{\mu\nu}+\frac{e^{\hat{\kappa}\chi}}{2}\partial_{\mu}\phi\partial_{\nu}\phi g^{\mu\nu}-V(\chi,\phi)\right],
\ee
which is obtained by switching to the Einstein frame from theories with nonminimal coupling to gravity. The background equations of motion in FLRW spacetime are
\be
H^2=\frac{\hat{\kappa}^2}{2}\rho, \qquad \dot{H}=-\frac{3}{4}\hat{\kappa}^2(\rho+p),
\ee
\be
\ddot{\chi}+3H\dot{\chi}-\frac{\hat{\kappa}}{2}e^{\hat{\kappa}\chi}\dot{\phi}^2+\frac{\partial V}{\partial\chi}=0, \qquad \ddot{\phi}+3H\dot{\phi}+\hat{\kappa}\dot{\chi}\dot{\phi}+e^{-\hat{\kappa}\chi}\frac{\partial V}{\partial\phi}=0
\ee
with
\be
\rho=\frac{1}{2}\left(\dot{\chi}^2+e^{\hat{\kappa}\chi}\dot{\phi}^2+2V\right), \qquad p=\frac{1}{2}\left(\dot{\chi}^2+e^{\hat{\kappa}\chi}\dot{\phi}^2-2V\right)
\ee
Then, we define the slow-roll parameters
\be
\alpha=\frac{\hat{\kappa}\dot{\chi}}{2H}, \qquad \lambda=\frac{\hat{\kappa}\dot{\phi}}{2H}e^{\frac{\hat{\kappa}}{2}\chi}
\ee
and assume that $\alpha , \lambda \ll 1$.
Rewriting the equations of motion in terms of the parameters $\alpha$ and $\lambda$ we find
\be
H^2(1-\alpha^2-\lambda^2)=\frac{\hat{\kappa}^2}{2}V, \qquad \dot{H}=-3H^2(\alpha^2+\lambda^2),
\ee
\be\label{eq:Vchi}
-v\beta_{\alpha}=\frac{\dot{\alpha}}{H}=-3\alpha(1-\alpha^2)-\lambda^2 (1+3\alpha)-\frac{\hat{\kappa}}{2}\frac{\partial_{\chi}V}{H^2},
\ee
\be\label{eq:Vphi}
-v\beta_{\lambda}=\frac{\dot{\lambda}}{H}=-3\lambda(1-\lambda^2)-\alpha\lambda (1-3\alpha)-\frac{\hat{\kappa}}{2}\frac{\partial_{\phi}V}{H^2}e^{-\frac{\hat{\kappa}}{2}\chi}.
\ee
Note that these equations are of the same type as~\eqref{eq:background1},~\eqref{eq:backgroundalpha} and~\eqref{eq:backgroundlambda}.

The equations analogous to~\eqref{eq:eqv},~\eqref{eq:eqchiphi},~\eqref{eq:eqH} are
\be\label{eq:eqvH2}
\beta_{\alpha}\frac{\partial v}{\partial\alpha}+\beta_{\lambda}\frac{\partial v}{\partial\lambda}=1-3(\alpha^2+\lambda^2)-v,\qquad \beta_{\alpha}\frac{\partial H}{\partial\alpha}+\beta_{\lambda}\frac{\partial H}{\partial\lambda}=3(\alpha^2+\lambda^2)\frac{H}{v},
\ee

\be\label{eq:eqchiphi2}
\beta_{\alpha}\frac{\partial \chi}{\partial\alpha}+\beta_{\lambda}\frac{\partial \chi}{\partial\lambda}=-\frac{2\alpha}{v\hat{\kappa}},\qquad \beta_{\alpha}\frac{\partial \phi}{\partial\alpha}+\beta_{\lambda}\frac{\partial \phi}{\partial\lambda}=-\frac{2\lambda e^{-\frac{\hat{\kappa}}{2}\chi}}{v\hat{\kappa}}.
\ee
We can see that they are the same, except for the second equation in~\eqref{eq:eqchiphi2}. In general, if we define $\alpha$ and $\lambda$ such that $\dot{H}=-3H^2(\alpha^2+\lambda^2)$, the equations~\eqref{eq:eqvH2} are always of this form, while the right-hand sides of~\eqref{eq:eqchiphi2} could be multiplied by some functions of the fields. Besides this difference, the solutions are uniquely determined once the type of beta function is chosen. An ansatz for the beta functions can be obtained by using the single-field results from~\cite{Anselmi:2021rye}, as we did in~\eqref{eq:betacond}. This is strongly dependent on the potential, rather than the field-space metric. Additional work in this direction might reveal a classification analogous to that of single-field potentials worked out in~\cite{Anselmi:2021rye}, which is beyond the aim of the present paper. 

To summarize, it is straightforward to apply the procedure presented here for different potentials and field-space metrics, provided that the slow-roll parameters are suitably defined.

%%%%%%%%%%%%%%%%%%%%%%%%%%%%%%%%%%%%%%%%%%%%%%%%%%%%%%%%%%%%%%%%%%%%%%%%%%%%%%%%%%%%%%
\sect{Running of slow-roll parameters}
\label{sec:running}

The slow-roll parameters $\alpha$ and $\lambda$ satisfy the equations~\eqref{eq:backgroundalpha} and~\eqref{eq:backgroundlambda}, which formally resemble those of running coupling constants in quantum field theory. In fact, in the previous section we have shown that the right-hand sides of~\eqref{eq:backgroundalpha} and~\eqref{eq:backgroundlambda} are power series in $\alpha$ and $\lambda$, and using the definition of beta function~\eqref{eq:vbeta} they can be written as
\be\label{eq:eqalpha}
\frac{\mathrm{d}\alpha(\ell)}{\mathrm{d}\ell}=-3\alpha\left[\alpha^2+(2-\varrho)\lambda^2\right]+\mathcal{O}(5),
\ee
\be\label{eq:eqlambda}
\frac{\mathrm{d}\lambda(\ell)}{\mathrm{d}\ell}=-3\lambda\left[\lambda^2+\left(2-\frac{1}{\varrho}\right)\alpha^2\right]+\mathcal{O}(5),
\ee
where we have introduced the rescaled conformal time $\eta$ and its logarithm
\be
\eta\equiv -k\tau, \qquad \ell\equiv \ln\eta.
\ee
Here $k$ is a parameter of mass-dimension $1$ which, in our analogy, plays the role of the renormalization scale and is later identified with the modulus of space momentum of cosmological perturbations. 

We choose the initial conditions at $\ell=0$, which means $\tau=-1/k$, and label them as $\alpha(0)=\alpha_k$ and $\lambda(0)=\lambda_k$. This introduces an implicit dependence on $k$, which was not present in the first place, since the equations~\eqref{eq:eqalpha} and ~\eqref{eq:eqlambda} are $k$-independent and so do their solutions. Such independence can be expressed as
\be\label{eq:kindependence}
\frac{\mathrm{d}\alpha}{\mathrm{d}\ln k}=0, \qquad \frac{\mathrm{d}\lambda}{\mathrm{d}\ln k}=0.
\ee
Again, this is in analogy with quantum field theory, where the coupling constants do not depend on the renormalization scale.
From equations~\eqref{eq:kindependence}, we can obtain the derivatives of $\alpha_k$ and $\lambda_k$ with respect to $\ln k$, which will be used in the next sections to derive the spectral indices. In fact, we can write
\be
\frac{\mathrm{d}\alpha}{\mathrm{d}\ln k}=\frac{\partial \alpha}{\partial\alpha_k}\frac{\mathrm{d} \alpha_k}{\mathrm{d}\ln k}+\frac{\partial \alpha}{\partial\lambda_k}\frac{\mathrm{d} \lambda_k}{\mathrm{d}\ln k}+\beta_{\alpha}(\alpha,\lambda)=0,
\ee
\be
\frac{\mathrm{d}\lambda}{\mathrm{d}\ln k}=\frac{\partial \lambda}{\partial\alpha_k}\frac{\mathrm{d} \alpha_k}{\mathrm{d}\ln k}+\frac{\partial \lambda}{\partial\lambda_k}\frac{\mathrm{d} \lambda_k}{\mathrm{d}\ln k}+\beta_{\lambda}(\alpha,\lambda)=0,
\ee
where in the last term of both equations we have used the fact that $\partial\ln\eta=\partial\ln|\tau|$. These equations hold for any $\ell$. In particular, we can set $\ell=0$ and since $\alpha_k$ and $\lambda_k$ do not depend on $\ell$ we get

\be\label{eq:lnkderivatives}
\frac{\mathrm{d}\alpha_k}{\mathrm{d}\ln k}=-\beta_{\alpha}(\alpha_k,\lambda_k), \qquad \frac{\mathrm{d}\lambda_k}{\mathrm{d}\ln k}=-\beta_{\lambda}(\alpha_k,\lambda_k).
\ee

Finally, the solutions of~\eqref{eq:eqalpha},~\eqref{eq:eqlambda} can be obtained perturbatively in $\alpha_k$ and $\lambda_k$. We find
\be
\begin{split}
\alpha(\ell)=&\alpha _k-3 \ell \alpha _k^3+3 \ell (\varrho-2) \alpha _k
   \lambda _k^2-\frac{3 \ell}{\varrho} \left[2+3 \varrho-\varrho^2+3 \ell\left(2-9 \varrho+4
   \varrho^2\right)\right] \alpha _k^3 \lambda _k^2\\
   &+\frac{3}{2} \ell \left[3 \ell \left(8-6
   \varrho+\varrho^2\right)+2 \left(-4+\varrho+\varrho^2\right)\right] \alpha _k \lambda _k^4+\frac{3}{2} \ell (9 \ell-4) \alpha _k^5+\mathcal{O}(6),
   \end{split}
\ee
\be
\begin{split}
\lambda(\ell)=&\lambda _k-3 \ell \lambda _k^3+\frac{3\ell}{\varrho} (1-2 \varrho) \alpha _k^2 \lambda _k-\frac{3 \ell}{\varrho}\left[3 \varrho-1+2 \varrho^2+3 \ell
   \left(4-9 \varrho+2 \varrho^2\right)\right] \alpha _k^2 \lambda _k^3\\
   &+\frac{3 \ell}{2 \varrho^2}
   \left[2+2 \varrho-8 \varrho^2+3 \ell \left(1-6 \varrho+8 \varrho^2\right)\right] \alpha _k^4\lambda _k+\frac{3}{2}
   \ell (9 \ell-4) \lambda _k^5+\mathcal{O}(6). 
\end{split}
\ee
These functions are useful to express quantities that are conserved in the superhorizon limit, where the dependence on $\ell$ drops and all the dependence on $k$ is encoded in $\alpha_k$ and $\lambda_k$, as shown at the end of the next section.

The equations~\eqref{eq:eqalpha} and~\eqref{eq:eqlambda} can also be solved exactly at each order of the beta functions. This is equivalent to resummation of the leading logs by powers of $\alpha_k^2\ell$, $\lambda_k^2\ell$ and $\alpha_k\lambda_k\ell$, pretty much like in quantum field theory, where the leading logs are resummed into the running coupling constants. However, for a general value of $\varrho$ it is not easy to find analytical solutions. Nevertheless, it is worth to notice that a change of variables can help, at least for the purpose of computing the number of e-folds before the end of inflation.
First, notice that if we use the variables
\be\label{eq:xiomega}
\xi\equiv x+y=\frac{\alpha^2+\varrho\lambda^2}{\sqrt{\alpha^2+\varrho^2\lambda^2}}, \qquad \omega\equiv \frac{x+\varrho y}{\sqrt{\varrho}}=\frac{\sqrt{\alpha^2+\varrho^2\lambda^2}}{\sqrt{\varrho}},
\ee
at the leading order the beta function of $\xi$ does not depend on $\omega$ and we find 
\be\label{eq:xieq}
\frac{\mathrm{d}\xi(\ell)}{\mathrm{d}\ell}=-3\xi^3+\mathcal{O}(5)
\ee
\be\label{eq:omegaeq}
\frac{\mathrm{d}\omega(\ell)}{\mathrm{d}\ell}=6\omega^3+3\omega\xi^2-\frac{6(\varrho+1)}{\sqrt{\varrho}}\omega^2\xi+\mathcal{O}(5).
\ee
At this order, the first equation can be solved analytically and with the initial condition $\xi(0)=\xi_k$ gives
\be
\xi(\ell)=\frac{\xi_k}{\sqrt{1+6\xi_k^2\ell}},
\ee
which is the same as for a single-field inflation with a cubic beta function~\cite{Anselmi:2021rye}. Then we define the variable $z\equiv\omega/\xi$ and take the ratio of~\eqref{eq:omegaeq} and~\eqref{eq:xieq} obtaining
\be\label{eq:zeq}
\xi\frac{\mathrm{d}z(\xi)}{\mathrm{d}\xi}=-2z+\frac{2(\varrho+1)}{\sqrt{\varrho}}z^2-2z^3.
\ee
In principle, knowing $z(\xi)$ and $\xi(\ell)$ can lead to $\omega(\ell)$. We can find a constant of motion from~\eqref{eq:zeq} by means of the separation of variables
\be
\ln\xi^2+\frac{1}{\varrho-1}\ln\left|\frac{z-\sqrt{\varrho}}{\sqrt{\varrho}z-1}\right|-\ln\left|\frac{\sqrt{\varrho}z-1}{z}\right|=\text{const.}
\ee
Again, depending on $\varrho$ an analytical solution for $z(\xi)$ might not exist. However, using the variables~\eqref{eq:xiomega} is enough to estimate the number of e-folds
\be
N(k)=\int_{t_k}^{t_f}\mathrm{d}tH(t),
\ee
where $t_k$ is the time when the modes with momentum $k$ exited the horizon and $t_f$ is the time when inflation ended. Recalling that $\dot{\xi}=-vH\frac{\mathrm{d}\xi}{\mathrm{d}\ell}$, at the leading order we have
\be\label{eq:Nk}
N(k)= \int_{\xi_k}^{\xi_f}\frac{\mathrm{d}\xi'}{\dot{\xi'}}H\simeq\int_{\xi_k}^{\xi_f}\frac{\mathrm{d}\xi'}{3\xi'^3}=\frac{1}{6\xi_k^2}+\mathcal{O}(0),
\ee
where $\xi_k=\xi(t_k)$ and $\xi_f=\xi(t_f)$. In terms of $\alpha_k$ and $\lambda_k$ \eqref{eq:Nk} gives
\be
N(k)\simeq\frac{\alpha_k^2+\varrho^2\lambda_k^2}{6(\alpha_k^2+\varrho\lambda_k^2)^2}.
\ee
In~\autoref{sec:perturbations} we discuss the relation between $N(k)$ and the tensor-to-scalar ratio.

%%%%%%%%%%%%%%%%%%%%%%%%%%%%%%%%%%%%%%%%%%%%%%%%%%%%%%%%%%%%%%%%%%%%%%%%%%%%%%%%%%%%%%
\sect{Perturbations}\label{sec:perturbations}
 In this section we study the perturbations of the metric and the scalar fields, derive their linearized equations of motion and obtain the power spectra from their solutions. The general strategy for these derivations works as follows. First, we parametrize the metric as
\begin{eqnarray}
g_{\mu \nu } &=&\text{diag}(1,-a^{2},-a^{2},-a^{2})-2a^{2}\left( u\delta
_{\mu }^{1}\delta _{\nu }^{1}-u\delta _{\mu }^{2}\delta _{\nu }^{2}+\tilde{u}\delta
_{\mu }^{1}\delta _{\nu }^{2}+\tilde{u}\delta _{\mu }^{2}\delta _{\nu }^{1}\right)
\notag \\
&&+2\text{diag}(\Phi ,a^{2}\Psi ,a^{2}\Psi ,a^{2}\Psi )-a\delta _{\mu
}^{0}\delta _{\nu }^{i}\partial _{i}B-a\delta _{\mu }^{i}\delta _{\nu
}^{0}\partial _{i}B, \label{mets}
\end{eqnarray}%
and the scalar fields as
\be
\chi(t,\mathbf{x})=\bar{\chi}(t)+\delta \chi(t,\mathbf{x}), \qquad \phi(t,\mathbf{x})=\bar{\phi}(t)+\delta \phi(t,\mathbf{x}),
\ee
where $\Psi, \Phi, B, u, \tilde{u}, \delta\chi,\delta\phi$ are perturbations while $\bar{\chi}$ and $\bar{\phi}$ denotes the background fields.

We work in the spatially-flat gauge, i.e. $\Psi=0$ and neglect vector perturbations, since they do not contribute in the superhorizon limit.

Then we perform the spatial Fourier transform, which for a generic perturbation $u(t,\mathbf{x})$ is written as $u_{\mathbf{k}}(t)$. To simplify the notation, we drop the subscript and for quadratic terms it is understood that $u^2$ means $u_{\mathbf{k}}u_{-\mathbf{k}}$, the same with $\dot{u}^2$, and the modulus of the space momentum is labeled as $k\equiv|\mathbf{k}|$. Moreover, in what follows we use the word “action" for the integral over time of the lagrangian density $\mathcal{L}_u$ of a perturbation $u$, and label it as $S_u$. This means that the full quadratic action is
\be
S=\int\frac{\mathrm{d}^3\mathbf{k}}{(2\pi)^3}S_u, \qquad S_u=\int\mathrm{d}t\mathcal{L}_u.
\ee

Finally, we apply a change of variables $u\rightarrow w$ such that the quadratic action in the rescaled conformal time $\eta$ reads
\be\label{eq:MSact}
S_w=\frac{1}{2}\int\mathrm{d}\eta\left(w'^2-w^2+\frac{2+\sigma}{\eta^2}w^2\right),
\ee
where $w'=\frac{\mathrm{d}w}{\mathrm{d}\eta}$ and $\sigma$ is a series in $\alpha$ and $\lambda$ that depends on the type of perturbation. This is also known as the Mukhanov-Sasaki action.

From~\eqref{eq:MSact} we get the corresponding equation of motion
\be\label{eq:MSeq}
w''+w-\frac{2+\sigma}{\eta^2}w=0,
\ee
which can be solved perturbatively by expanding its solution $w(\eta)$ in powers of $\alpha_k$ and $\lambda_k$
\be
w(\eta)=\sum_{n,m=0}^{\infty}w_{nm}(\eta)\alpha_k^n\lambda_k^m,
\ee
where $w_{nm}(\eta)$ are functions to be determined. 

In the following we are going to discuss solutions of \eqref{eq:MSeq} in the superhorizon limit, i.e. for  $k/(aH)\to 0$. It is useful to rewrite the limit in terms of $\eta$ 
adopting the definition of $v$ from~\eqref{eq:vbeta}:
\be
\frac{k}{aH}=\eta v=\eta\left[1+\mathcal{O}(2)\right].
\ee
Then it is clear that the superhorizon limit $k/(aH)\rightarrow 0$ corresponds to $\eta\rightarrow 0$. Analogously, the opposite limit $k/(aH)\rightarrow \infty$ corresponds to
$\eta\rightarrow \infty$, which is where we fix the Bunch-Davies boundary condition
\be\label{eq:bd}
w(\eta)\rightarrow\frac{e^{i\eta}}{\sqrt{2}}, \qquad \text{for} \qquad \eta\rightarrow\infty,
\ee 
since $\eta\rightarrow\infty$ is the infinite past. In terms of $w_{nm}$ the condition~\eqref{eq:bd} reads
\be
w_{00}(\eta)\rightarrow\frac{e^{i\eta}}{\sqrt{2}}, \qquad w_{nm}(\eta)\rightarrow 0 \qquad \text{for} \qquad \eta\rightarrow\infty
\ee
for every other combination of $n$ and $m$. Given $n$ and $m$ the equation for $w_{nm}$ takes the form
\be\label{eq:eqwnm}
w''_{nm}+w_{nm}-\frac{2}{\eta^2}w_{nm}=\frac{f_{nm}(\eta)}{\eta^2},
\ee
where $f_{nm}(\eta)$ are functions determined once the $w_{nm}$ of the previous orders are derived. Then the solution can be written as~\cite{Anselmi:2020shx}
\be
w_{nm}(\eta)=\int_{\eta}^{\infty}\mathrm{d}\eta'\frac{f_{nm}(\eta')}{\eta\eta^{\prime 3}}\left[(\eta-\eta')\cos (\eta-\eta')-(1+\eta\eta')\sin (\eta-\eta')\right].
\ee

In general, tensor and scalar perturbations require separated discussions. However, we report here a set of functions that are useful to write the solutions of~\eqref{eq:eqwnm} in a more compact form. Following the notation of~\cite{Anselmi:2021rye} we define
\begin{eqnarray}
W_{0} &=&\frac{i(1-i\eta )}{\eta \sqrt{2}}\mathrm{e}^{i\eta },\qquad W_{2}=3%
\left[ \text{Ei}(2i\eta )-i\pi \right] W_{0}^{\ast }+\frac{6W_{0}}{(1-i\eta )%
},  \notag \\
W_{3} &=&\left[ 6(\ln \eta +\tilde{\gamma}_{M})^{2}+24i\eta
F_{2,2,2}^{1,1,1}\left( 2i\eta \right) +\pi ^{2}\right] W_{0}^{\ast }+\frac{%
24W_{0}}{(1-i\eta )}-4(\ln \eta +1)W_{2},  \label{Wi} \\
W_{4} &=&-\frac{16W_{0}}{1+\eta ^{2}}+\frac{2(13+i\eta )W_{2}}{9(1+i\eta )}+%
\frac{W_{3}}{3}+4G_{2,3}^{3,1}\left( -2i\eta \left\vert _{0,0,0}^{\
0,1}\right. \right) W_{0},  \notag
\end{eqnarray}%
where $\tilde{\gamma}_M=\gamma_M-\frac{i\pi}{2}=\gamma_{E}+\ln 2-\frac{i\pi}{2}$, $\gamma_E$ being the Euler-Mascaroni constant, while Ei denotes the exponential-integral function, $F_{b_{1},\cdots
,b_{q}}^{a_{1},\cdots ,a_{p}}(z)$ is the generalized hypergeometric function 
$_{p}F_{q}(\{a_{1},\cdots ,a_{p}\},\{b_{1},\cdots ,b_{q}\};z)$ and $%
G_{p,q}^{m,n}$ is the Meijer G function.

Before we proceed it is useful to highlight some properties of the equation~\eqref{eq:MSeq} and its solution. In~\cite{Anselmi:2020shx}, it was proved that the solution $w(\eta)$ can be decomposed as
\be\label{eq:wqy}
\eta w=Q(\ln \eta)+Y(\eta),
\ee
where $Q$ is a power series in $\ln\eta$ while $Y$ is a power series in $\eta$ and $\ln\eta$ such that $Y\rightarrow 0$ term by term for $\eta\rightarrow 0$. Since we are interested in the correlation functions in the superhorizon limit, the quantity $Q$ is the most relevant one. It is possible to find an equation for $Q$ by inserting the decomposition~\eqref{eq:wqy} in the equation~\eqref{eq:MSeq}, which gives
\be
\frac{\mathrm{d}^2Q}{\mathrm{d}\ln\eta^2}-3\frac{\mathrm{d}Q}{\mathrm{d}\ln\eta}-\sigma Q=\eta ^{2}\frac{\mathrm{d}^2Y}{\mathrm{d}\eta^2}-2\eta \frac{\mathrm{d}Y}{\mathrm{d}\eta}-\sigma Y-\eta ^{2}(Y+Q).
\ee
The right-hand side of the equation above goes to zero for $\eta\rightarrow 0$, then both sides are zero independently, and for $Q$ we get
\be\label{eq:Qetaeq}
\left(\frac{\mathrm{d}}{\mathrm{d}\ln\eta}-3\right)\frac{\mathrm{d}Q}{\mathrm{d}\ln\eta}=\sigma Q.
\ee
Furthermore, we can perturbatively invert the operator in the parenthesis to get rid of solutions that goes to zero for $\eta\rightarrow 0$, which by definition do not belong to $Q$, i.e.
\be
\frac{\mathrm{d}Q}{\mathrm{d}\ln \eta }=-\frac{1}{3}\frac{1}{1-\frac{1}{3}%
\frac{\mathrm{d}}{\mathrm{d}\ln \eta }}\left(\sigma Q\right),
\ee
where the operator on the right-hand side must be understood as the perturbative series
\be
\frac{1}{1-\frac{1}{3}%
\frac{\mathrm{d}}{\mathrm{d}\ln \eta }}F(\ln\eta)=F(\ln\eta)+\sum_{n=1}^{\infty}\frac{1}{3^n}\frac{\mathrm{d}^nF(\ln\eta)}{\mathrm{d}\ln^n\eta},
\ee
$F$ being any function of $\ln\eta$.

Finally, if we fix a reference scale $k$, we can view $Q(\ln\eta)$ as a function $\tilde{Q}(\alpha,\lambda,\alpha_k,\lambda_k)$ and write
\be
\frac{\mathrm{d}}{\mathrm{d}\ln\eta}=\beta_{\alpha}\frac{\partial}{\partial\alpha}+\beta_{\lambda}\frac{\partial}{\partial\lambda}\equiv D.
\ee
Then $\tilde{Q}$ satisfies
\be\label{eq:eqQtilde}
D\tilde{Q}=-\frac{1}{3}\frac{1}{1-\frac{1}{3}D}\left(\sigma \tilde{Q}\right).
\ee
The solution of\eqref{eq:eqQtilde} can be parametrized as
\be\label{eq:QJ}
\tilde{Q}(\alpha,\lambda,\alpha_k,\lambda_k)=\tilde{Q}_{0}(\alpha_k,\lambda_k)\frac{J(\alpha,\lambda)}{J(\alpha_k,\lambda_k)},
\ee
where $\tilde{Q}_{0}(\alpha_k,\lambda_k)\equiv Q(0)$, and $J$ satisfies~\eqref{eq:eqQtilde}.
The advantage of introducing $J$ is that it allows us to find perturbations that are conserved in the superhorizon limit without spoiling their gauge invariance. In fact, we can define the perturbation
\be\label{eq:wC}
w^{\text{C}}(\eta)\equiv\frac{C}{k^{3/2}}\frac{\eta w(\eta)}{J(\alpha,\lambda)}=\frac{C}{k^{3/2}}\left[\frac{\tilde{Q}_0(\alpha_k,\lambda_k)}{J(\alpha_k,\lambda_k)}+\frac{Y(\eta)}{J(\alpha,\lambda)}\right],
\ee
where the factor $k^{3/2}$ is introduced for dimensional reasons and $C$ is an arbitrary dimensionless constant. It is easy to see that the perturbation~\eqref{eq:wC} is conserved in the superhorizon limit by construction. Indeed,
\be
w^{\text{C}}(\eta)\simeq\frac{C}{k^{3/2}}\frac{\tilde{Q}_0(\alpha_k,\lambda_k)}{J(\alpha_k,\lambda_k)}, \qquad \text{for} \ \eta\to 0.
\ee
Finally, since $J$ is a function of background quantities only, if $\eta w$ is gauge invariant, also $w^{\text{C}}$ is gauge invariant.

Here we have shown the derivation of the equation~\eqref{eq:eqQtilde} for a single perturbation $w(\eta)$. However, since in this paper we consider a double-field model, a similar equation can be derived for scalar perturbations, where $w$ would be a 2-component vector while $\sigma$ a $2\times 2$ matrix.

In the following two subsections we adopt the strategy described here for tensor and scalar perturbations.

%%%%%%%%%%%%%%%%%%%%%%%%%%%%%%%%%%%%%%%%%%%%%%%%%%%%%%%%%%%%%%%%%%%%%%%%%%%%%%%%%%%%%%
\subsection{Tensor perturbations}
Tensor perturbations are parametrized by $u$ and $\tilde{u}$, which are gauge invariant. The quadratic action for $\tilde{u}$ is identical to that of $u$, so we derive only the latter and account for $\tilde{u}$ at the end. To the quadratic order in $u$, the action is
\be
S_{u}=\frac{1}{\kappa^2}\int\mathrm{d}ta^3\left(\dot{u}^2-\frac{k^2}{a^2}u^2\right).
\ee
To obtain the action in the Mukhanov-Sasaki form~\eqref{eq:MSact}, first we perform the change of variables
\be\label{eq:uredefinition}
u=\frac{\kappa}{\sqrt{2k}a}w_{\text{t}},
\ee
then switch to the rescaled conformal time $\eta$. Finally, using the definitions~\eqref{eq:vbeta} we find
\be
S_{w_\text{t}}=\frac{1}{2}\int\mathrm{d}\eta\left(w_{\text{t}}'^2-w_{\text{t}}^2-\frac{2+\sigma_{\text{t}}}{\eta^2}w_{\text{t}}^2\right),
\ee
where 
\be
\sigma_{\text{t}}=\frac{2(1-v^2)-3(\alpha^2+\lambda^2)}{v^2},
\ee
and using~\eqref{eq:vseries} we get

\be
\begin{split}
\sigma_{\text{t}}&= 9 \left(\alpha ^2+\lambda ^2\right)+108 \left[\alpha ^4+\lambda ^4+\frac{2}{3} \alpha ^2 \lambda ^2
   \left(5-\frac{1}{\varrho }-\varrho \right)\right]+1683 \left(\alpha ^6+\lambda ^6\right)\\
   &+9 \alpha ^4 \lambda ^2 \left(1217+\frac{40}{\varrho ^2}-\frac{388}{\varrho }-308 \varrho
   \right)+9 \alpha ^2 \lambda ^4 \left(1217-\frac{308}{\varrho }-388 \varrho +40 \varrho ^2\right)+\mathcal{O}(8).
   \end{split}
\ee
Writing
\be\label{eq:wtexpansion}
w_{\text{t}}=\sum_{n,m}w_{nm}^{(\text{t})}\alpha_k^n\lambda_k^m
\ee 
and solving the equation~\eqref{eq:eqwnm} for each $w_{nm}^{(\text{t})}$, up to order 4 we obtain
\be
w_{00}^{(\text{t})}=W_0, \quad w_{02}^{(\text{t})}=w_{20}^{(\text{t})}=W_2, 
\ee
\be
w_{01}^{(\text{t})}=w_{10}^{(\text{t})}=w_{11}^{(\text{t})}=w_{21}^{(\text{t})}=w_{12}^{(\text{t})}=w_{03}^{(\text{t})}=w_{30}^{(\text{t})}=0, 
\ee
\be
 w_{22}^{\text{(t)}}=5W_2-\frac{3}{2\varrho}(1-3\varrho+\varrho^2)W_3+\frac{9}{2}W_4\quad w_{40}^{(\text{t})}=w_{04}^{(\text{t})}=\frac{5}{3}W_2+\frac{3}{4}W_3+\frac{9}{4}W_4.
\ee

As explained in the previous subsection, we use the decomposition~\eqref{eq:wqy} for $w_{\text{t}}$ 
\be
\eta w_{\text{t}}=Q_{\text{t}}(\ln \eta)+Y_{\text{t}}(\eta).
\label{eq:Q-Y}
\ee
Then $\tilde{Q}_{\text{t}}$ satisfies

\be\label{eq:eqQtildet}
D\tilde{Q}_{\text{t}}=-\frac{1}{3}\frac{1}{1-\frac{1}{3}D}\left(\sigma_{\text{t}} \tilde{Q}_{\text{t}}\right)
\ee
and following~\eqref{eq:QJ} we parametrize $\tilde{Q}_{\text{t}}$  as
\be\label{eq:QJt}
\tilde{Q}_{\text{t}}(\alpha,\lambda,\alpha_k,\lambda_k)=\tilde{Q}_{0\text{t}}(\alpha_k,\lambda_k)\frac{J_\text{t}(\alpha,\lambda)}{J_\text{t}(\alpha_k,\lambda_k)},
\ee
where $\tilde{Q}_{0\text{t}}(\alpha_k,\lambda_k)\equiv Q_{\text{t}}(0)$. Then the conserved perturbation is
\be\label{eq:tensorRG}
w^{\text{C}}_{\text{t}}(\eta)\equiv\frac{C_{\text{t}}}{k^{3/2}}\frac{\eta w_{\text{t}}(\eta)}{J_{\text{t}}(\alpha,\lambda)},\qquad C_{\text{t}}=\frac{m_{\chi}\hat{\kappa}}{2\sqrt{3}},
\ee
where the constant $C_{\text{t}}$ is chosen so that \be
w^{\text{C}}(\eta)=u(t(\eta)).
\ee
In fact, in the case of tensor perturbations there is only one independent gauge-invariant perturbation $u$, which is also conserved in the superhorizon limit. Therefore, the perturbation $w^{\text{C}}$ that we just build must coincide with $u$, up to an arbitrary constant. To prove this we compare~\eqref{eq:uredefinition} with~\eqref{eq:tensorRG} and show that 
\be\label{eq:HvJconst}
\frac{m_{\chi}}{3HvJ_{\text{t}}}=\text{constant}.
\ee
Using~\eqref{eq:eqv} and~\eqref{eq:eqH} it is easy to see that $\frac{m_{\chi}}{3Hv}$ satisfies~\eqref{eq:Qetaeq}, hence the relation~\eqref{eq:HvJconst}. Setting the arbitrary constant to one we get
\be\label{eq:JHv}
J_{\text{t}}(\alpha,\lambda)=\frac{m_{\chi}}{3Hv}=\frac{\sqrt{\alpha^2+\varrho\lambda^2}}{\tilde{H}v}.
\ee
Therefore, the function $J_{\text{t}}$ can be obtained either by solving~\eqref{eq:eqQtildet}, or by means of~\eqref{eq:JHv}. Finally, $\tilde{Q}_{0\text{t}}$ is derived by comparing~\eqref{eq:QJ} with~\eqref{eq:wtexpansion} at $\ln\eta=0$ order by order. The results are
\be
\frac{J_{\text{t}}(\alpha,\lambda)}{\sqrt{\alpha^2+\varrho \lambda^2}}=1+\frac{7}{2}\left(\alpha^2+\lambda^2\right)+\frac{223}{8}\left(\alpha^4+\lambda^4\right)+\frac{\alpha^2\lambda^2}{4}\left(355-66\zeta\right)+\mathcal{O}(6),
\ee
\be
\begin{split}
\tilde{Q}&_{0\text{t}}(\alpha_k,\lambda_k)=\frac{i}{\sqrt{2}}\left[1-3(\alpha_k^2+\lambda_k^2)(\tilde{\gamma}_M-2)+\frac{3(\alpha_k^4+\lambda_k^4)}{4} \left(24+\pi ^2-6 \tilde{\gamma }_M \left(2+\tilde{\gamma }_M\right)\right)\right.\\
&\left. +\frac{3\alpha_k^2\lambda_k^2}{2}\left(6 \left(4+2 \tilde{\gamma} _M-3 \tilde{\gamma} _M^2\right)-\pi ^2-\zeta  \left(6 \left(2-\tilde{\gamma} _M\right) \tilde{\gamma} _M-\pi ^2\right)\right)\right]+\mathcal{O}(6).
\end{split}
\ee

The two-point function is derived as usual by canonically quantizing $w^{\text{C}}_{\text{t}}$. We introduce the operator
\be
\hat{w}^{\text{C}}_{\text{t}\mathbf{k}}(\eta )=w_{\text{t}\mathbf{k}}^{\text{C}}(\eta )\hat{a}_{\mathbf{k}}+w_{\text{t}-%
\mathbf{k}}^{\text{C}\ast }(\eta )\hat{a}_{-\mathbf{k}}^{\dagger },
\ee
where $\hat{a}_{\mathbf{k}}^{\dagger }$ and $\hat{a}_{\mathbf{k}}$ are
creation and annihilation operators which satisfy 
\be
[\hat{a}_{\mathbf{k}},\hat{a}%
_{\mathbf{k}^{\prime }}^{\dagger }]=(2\pi )^{3}\delta ^{(3)}(\mathbf{k}-%
\mathbf{k}^{\prime }).
\ee
Then the power spectrum $\mathcal{P}_{w_{\text{t}}^{\text{C}}}$ of the perturbations 
$w^{\text{C}}_{\text{t}}$ is defined by the two-point function%
\begin{equation}
\langle \hat{w}_{\text{t}\mathbf{k}}^{\text{C}}(\eta )\hat{w}_{\text{t}\mathbf{k}^{\prime
}}^{\text{C}}(\eta )\rangle =(2\pi )^{3}\delta ^{(3)}(\mathbf{k}+\mathbf{k}%
^{\prime })\frac{2\pi ^{2}}{k^{3}}\mathcal{P}_{w_{\text{t}}^{\text{C}}},  \label{2pt}
\end{equation}%
which gives the tensor power spectrum
\be
\mathcal{P}_{\text{t}}\equiv16\mathcal{P}_{w_{\text{t}}^{\text{C}}}=16\frac{k^{3}}{2\pi ^{2}}\left\vert w_{\text{t}\mathbf{k}}^{\text{C}%
}\right\vert ^{2},
\ee
where the factor 16 comes form the summation over tensor polarizations.
In the superhorizon limit, we obtain for the tensor power spectrum
\begin{equation}
\mathcal{P}_\text{t}\simeq 16\frac{|C_{\text{t}}|^{2}}{2\pi ^{2}}\left\vert \frac{\tilde{Q}_{0\text{t}}(\alpha _{k},\lambda_k)}{J_{\text{t}}(\alpha _{k},\lambda_k)}\right\vert ^{2}.  \label{spectrum}
\end{equation}%
The result to the next-to-next-to-leading order is

\be
\label{eq:Pt}
\begin{split}
\mathcal{P}_{\text{t}}=&\frac{2m_{\chi}^2\kappa^2}{9 \pi^2(\alpha_k^2+\varrho \lambda_k^2)}\Bigg\{1+(\alpha_k^2+\lambda_k^2)(5-6\gamma_M)-(\alpha_k^4+\lambda_k^4)(31+12\gamma_M-6\pi^2)\\
&-\alpha_k^2\lambda_k^2\left[128-15 \pi ^2-12 \gamma _M \left(4-3 \gamma _M\right)-\frac{3}{2}\zeta\left(22-\pi ^2-12 (2-\text{$\gamma_M$}) \text{$\gamma_M$}\right)\right]+\mathcal{O}(6)\Bigg\}.
\end{split}
\ee

Finally, we derive the spectral index, which is defined as 
\be
n_{\text{t}}\equiv\frac{\mathrm{d}\ln\mathcal{P}_{\text{t}}}{\mathrm{d}\ln k}=\frac{\mathrm{d}\alpha_k}{\mathrm{d}\ln k}\frac{\partial\ln\mathcal{P}_{\mathrm{t}}}{\partial\alpha_k}+\frac{\mathrm{d}\lambda_k}{\mathrm{d}\ln k}\frac{\partial\ln\mathcal{P}_{\mathrm{t}}}{\partial\lambda_k}.
\ee
Then using the equations~\eqref{eq:lnkderivatives} we get

\be
n_{\text{t}}=-\beta_{\alpha}(\alpha_k,\lambda_k)\frac{\partial\ln\mathcal{P}_{\text{t}}}{\partial\alpha_k}-\beta_{\lambda}(\alpha_k,\lambda_k)\frac{\partial\ln\mathcal{P}_{\text{t}}}{\partial\lambda_k}.
\ee
To the NNL order this gives
\be
n_{\text{t}}=-6(\alpha_k^2+\lambda_k^2)+18(1-2 \gamma_M)(\alpha_k^4+\lambda_k^4)
+36\alpha_k^2\lambda_k^2\left[3-4\gamma_M-\zeta(1-\gamma_M)\right]+\mathcal{O}(6).
\ee

This method is crucial in the case of scalar perturbations, where the usual adiabatic and isocurvature perturbations are not conserved and it is not trivial to find combinations that are both gauge-invariant and conserved in the superhorizon limit 
.

%%%%%%%%%%%%%%%%%%%%%%%%%%%%%%%%%%%%%%%%%%%%%%%%%%%%%%%%%%%%%%%%%%%%%%%%%%%%%%%%%%%%%%
\subsection{Scalar perturbations}
\label{subsec:scalarpert}
In the spatially-flat gauge, the relevant fields for the study of scalar perturbations are $\Phi$, $B$, $\delta\chi$ and $\delta\phi$. The first two are auxiliary fields and after integrating them out the quadratic action is
\be
S_{\delta\chi\delta\phi}=\frac{1}{2}\int\mathrm{d}ta^3\left[\dot{\delta\chi}^2-\Omega_{\chi}\delta\chi^2+\dot{\delta\phi}^2-\Omega_{\phi}\delta\phi^2+2\Omega_{\chi\phi}\delta\chi\delta\phi\right],
\ee
where
\be
\Omega_{\chi}\equiv m_{\chi}^2+\frac{k^2}{a^2}+6H^2\left[2v\alpha\beta_{\alpha}-3\alpha^2\left(1-\alpha^2-\lambda^2\right)\right],
\ee

\be
\Omega_{\phi}\equiv m_{\phi}^2+\frac{k^2}{a^2}+6H^2\left[2v\lambda\beta_{\lambda}-3\lambda^2\left(1-\alpha^2-\lambda^2\right)\right],
\ee

\be
\Omega_{\chi\phi}\equiv 6H^2\left[v\left(\alpha\beta_{\lambda}+\lambda\beta_{\alpha}\right)-3\alpha\lambda\left(1-\alpha^2-\lambda^2\right)\right].
\ee
Then, we perform the change of variables
\be
\delta\chi=\frac{U}{a\sqrt{k}},\qquad \delta\phi=\frac{V}{a\sqrt{k}}
\ee
and move to the rescaled conformal time $\eta$. In this way, we obtain the action in the Mukhanov-Sasaki form
\be
S_{w_\text{s}}=\frac{1}{2}\int\mathrm{d}\eta\left[w^{\prime  \text{T}}_{\text{s}}w'_{\text{s}}-\left(1-\frac{2}{\eta^2}\right)w^{\text{T}}_{\text{s}}w_{\text{s}}+\frac{1}{\eta^2}w^{\text{T}}_{\text{s}}\Sigma w_{\text{s}}\right], \qquad w_{\text{s}}\equiv\begin{pmatrix}
U\\
V
\end{pmatrix},
\ee
where
\be
\Sigma=\begin{pmatrix}
\sigma_U& \sigma_{UV}\\
\sigma_{UV}&\sigma_V
\end{pmatrix},
\ee
\be
\sigma_U=\sigma_{\text{t}}-\frac{m_{\chi}^2}{H^2v^2}-\frac{6\alpha}{v^2}\left[ 2\beta_{\alpha}v-3\alpha\left(1-\alpha^2-\lambda^2\right)\right],
\ee
\be
\sigma_V=\sigma_{\text{t}}-\frac{m_{\phi}^2}{H^2v^2}-\frac{6\lambda}{v^2}\left[2\beta_{\lambda}v-3\lambda\left(1-\alpha^2-\lambda^2\right)\right],
\ee
\be
\sigma_{UV}=-\frac{6}{v^2}\left[v\left(\lambda\beta_{\alpha}+\alpha\beta_{\lambda}\right)-3\alpha\lambda\left(1-\alpha^2-\lambda^2\right)\right].
\ee
At the leading order the matrix $\Sigma$ reads
\be
\Sigma=18\begin{pmatrix}
2\alpha^2+\frac{1}{2}(1-\rho)\lambda^2& \alpha\lambda\\
\alpha\lambda&2\lambda^2+\frac{1}{2}\left(1-\frac{1}{\rho}\right)\alpha^2
\end{pmatrix}+\mathcal{O}(4).
\ee
Following the same approach of the previous subsection, we write
\be\label{eq:wsexpansion}
w_{\text{s}}=\sum_{n,m}w_{nm}^{(\text{s})}\alpha_k^n\lambda_k^m
\ee
and solve the equations up to the order $\mathcal{O}(4)$ imposing the Bunch-Davies condition  
\be
w_{\text{s}}\rightarrow\frac{e^{i\eta}}{\sqrt{2}}\begin{pmatrix}
    1\\1
\end{pmatrix}, \qquad \eta\rightarrow\infty.
\ee
We find

\be
w_{00}^{(\text{s})}=W_0\begin{pmatrix} 1\\ 1\end{pmatrix}, \quad  w_{11}^{(\text{s})}=2W_2\begin{pmatrix} 1\\ 1\end{pmatrix},\quad w_{02}^{(\text{s})}=W_2\begin{pmatrix} 2\\ 1-\varrho\end{pmatrix}, \quad w_{20}^{(\text{s})}=W_2\begin{pmatrix} 2\\ 1-1/\varrho\end{pmatrix},
\ee

\be
w_{22}^{(\text{s})}=W_2\begin{pmatrix} 19\varrho-17\\[1ex]
\frac{19}{\varrho}-17\end{pmatrix}+\frac{3}{2}W_3\begin{pmatrix} 3-\frac{1}{\varrho}-2\varrho\\[1ex] 3-\varrho-\frac{2}{\varrho}\end{pmatrix}+9W_4\begin{pmatrix} \varrho-2\\[1ex] \frac{1}{\varrho}-2\end{pmatrix},
\ee

\be
w_{40}^{(\text{s})}=\frac{1}{2}W_2\begin{pmatrix} -3\\[1ex]5+\frac{8}{\varrho}-\frac{3}{\varrho^2}\end{pmatrix}+\frac{3}{4}W_3\begin{pmatrix} 0\\[1ex]\frac{\varrho^2-1}{\varrho^2}\end{pmatrix}+9W_4\begin{pmatrix} 1\\[1ex]\frac{(\varrho-1)^2}{4\varrho^2}
\end{pmatrix},
\ee

\be
w_{04}^{(\text{s})}=\frac{1}{2}W_2\begin{pmatrix} 5+8\varrho-3\varrho^2\\[1ex]-3\end{pmatrix}+\frac{3}{4}W_3\begin{pmatrix} 1-\varrho^2\\[1ex]0\end{pmatrix}+9W_4\begin{pmatrix} \frac{(\varrho-1)^2}{4}\\[1ex]1
\end{pmatrix},
\ee

\be
w_{01}^{(\text{s})}=w_{10}^{(\text{s})}=w_{21}^{(\text{s})}=w_{12}^{(\text{s})}=w_{30}^{(\text{s})}=w_{03}^{(\text{s})}=0.
\ee
The field operators are defined as
\be
\hat{w}_{\text{s}\mathbf{k}}=w_{\text{s}\mathbf{k}}\hat{a}_{\mathbf{k}}+w^*_{\text{s}-\mathbf{k}}\hat{a}^{\dagger}_{-\mathbf{k}}, \qquad \hat{a}_{\mathbf{k}}\equiv\begin{pmatrix}
\hat{a}_{U\mathbf{k}}\\
\hat{a}_{V\mathbf{k}}
\end{pmatrix}
\ee
with
\be
\left[\hat{a}_{\mathbf{k}},\hat{a}^{\dagger}_{\mathbf{k}'}\right]=(2\pi)^3\delta^{(3)}(\mathbf{k}-\mathbf{k}')\mathds{1},\qquad \left[\hat{a}_{\mathbf{k}},\hat{a}_{\mathbf{k}'}\right]=\left[\hat{a}^{\dagger}_{\mathbf{k}},\hat{a}^{\dagger}_{\mathbf{k}'}\right]=0,
\ee
where $\hat{a}_{U\mathbf{k}}$ and $\hat{a}_{V\mathbf{k}}$ are the creation operators of $U$ and $V$, respectively. At this point, there is an important difference between the tensor and scalar perturbations. While in the tensor case we have only one independent gauge-invariant variable $u$, in this case we have two independent ways of building gauge-invariant variables. Indeed, given a scalar quantity $\varphi=\bar{\varphi}+\delta\varphi$, where $\bar{\varphi}$ is the background and $\delta\varphi$ its perturbation, we can always build a gauge-invariant variable as
\be
\mathcal{R}_{\varphi}=\Psi+H\frac{\delta\varphi}{\dot{\bar{\varphi}}}.
\ee
Since in our model we have two independent scalars, then in the spatially-flat gauge, we have
\be\label{eq:RchiRphi}
\mathcal{R}_{\chi}=H\frac{\delta\chi}{\dot{\bar{\chi}}}=\frac{\hat{\kappa}Hv\eta U}{2 k^{3/2}\alpha}, \qquad \mathcal{R}_{\phi}=H\frac{\delta\phi}{\dot{\bar{\phi}}}=\frac{\hat{\kappa}Hv\eta V}{2 k^{3/2}\lambda}.
\ee
Moreover, any independent pair of linear combinations
\be\label{eq:Rcombinations}
A \mathcal{R}_{\chi}+B\mathcal{R}_{\phi}, \qquad C \mathcal{R}_{\chi}+D\mathcal{R}_{\phi},
\ee
where $A, \ B, \ C$ and $D$ are functions of background quantities, is also gauge invariant, and in general they are not conserved in the superhorizon limit. For example, the perturbations~\eqref{eq:RchiRphi} in the superhorizon limit are

\be
\begin{split}
\mathcal{R}_{\chi}=\frac{i}{\alpha_k\sqrt{\alpha_k^2+\varrho \lambda_k^2}}\frac{\hat{\kappa}m_{\chi}}{k^{3/2}6\sqrt{2}}&\left[1+\alpha_k^2\left(\frac{17}{2}-6\tilde{\gamma}_M\right)+\lambda_k^2\left(\frac{5}{2}+3\tilde{\gamma}_M(\varrho-1)-6\varrho+6\ell\right)\right.\\
&-6\alpha_k\lambda_k\left(\tilde{\gamma}_M-2+\ell\right)\bigg]+\mathcal{O}(4)
\end{split}
\ee
and analogous expression for $\mathcal{R}_{\phi}$ obtained from $\mathcal{R}_{\chi}$ by means of the exchanges $\alpha_k\leftrightarrow\lambda_k$ and $m_{\chi}\leftrightarrow m_{\phi}$. Both these perturbations are not conserved\footnote{Note that for $\lambda_k=0$ (single field case) or $\lambda_k=\alpha_k$ and $\varrho=1$ (two copies of the same field), $\mathcal{R}_{\chi}$ is conserved.} since they depend explicitly on $\ell$.

Another example, typically used in the context of double-field inflation, is given by the so-called adiabatic and isocurvature perturbations, which are of the form~\eqref{eq:Rcombinations} with
\be
A=\frac{\dot{\bar{\chi}}^2}{\dot{\bar{\chi}}^2+\dot{\bar{\phi}}^2}, \qquad B=\frac{\dot{\bar{\phi}}^2}{\dot{\bar{\chi}}^2+\dot{\bar{\phi}}^2}, \qquad C=-D=\frac{\dot{\bar{\chi}}\dot{\bar{\phi}}}{\dot{\bar{\chi}}^2+\dot{\bar{\phi}}^2}.
\ee
In the spatially-flat gauge they are
\be
\mathcal{R}_{\text{ad}}=H\frac{\dot{\bar{\chi}}\delta\chi+\dot{\bar{\phi}}\delta\phi}{\dot{\bar{\chi}}^2+\dot{\bar{\phi}}^2}, \qquad \mathcal{R}_{\text{iso}}=H\frac{\dot{\bar{\phi}}\delta\chi-\dot{\bar{\chi}}\delta\phi}{\dot{\bar{\chi}}^2+\dot{\bar{\phi}}^2},
\ee
respectively. Also these perturbations are not conserved in the superhorizon limit. Therefore, we would like to find alternative combinations of the form~\eqref{eq:Rcombinations} that are conserved.

In order to find such quantities, we proceed in analogy with the tensor perturbations and decompose $w_{\text{s}}$ as
\be\label{eq:eqQs}
\eta w_{\text{s}}=Q_{\text{s}}(\ln\eta)+Y_{\text{s}}(\eta),
\ee
where now $Q_{\text{s}}$ and $Y_{\text{s}}$ are 2-component vectors. Then $Q_{\text{s}}$ satisfies the analog of~\eqref{eq:eqQtilde}, which reads
\be\label{eq:eqQDs}
DQ_{\text{s}}=-\frac{1}{3}\frac{1}{1-\frac{1}{3}D}\Sigma Q_{\text{s}}.
\ee
We parametrize the solutions as
\be
\tilde{Q}_{\text{s}}(\alpha,\lambda,\alpha_k,\lambda_k)=\mathcal{J}(\alpha,\lambda)\mathcal{J}^{-1}(\alpha_k,\lambda_k)\tilde{Q}_{0\text{s}}(\alpha_k,\lambda_k),
\ee
where $\mathcal{J}$ is a $2\times 2$ matrix
and $\tilde{Q}_{0\text{s}}$ is a constant vector. The matrix $\mathcal{J}$ can be found by solving~\eqref{eq:eqQDs}, while the vector $\tilde{Q}_{0\text{s}}$ is determined by comparison with $\eta w_{\text{s}}$.  We obtain
\be
\mathcal{J}(\alpha,\lambda)=\begin{pmatrix}
\alpha Z J_1& -\frac{\sqrt{\varrho}\lambda^2\alpha}{Z^3}J_2 \\[2ex]
\lambda Z J_1& \frac{\lambda\alpha^2}{\sqrt{\varrho}Z^3}J_2
\end{pmatrix}, \qquad Z\equiv \sqrt{\alpha^2+\varrho\lambda^2},
\ee

\be
\tilde{Q}_{0\text{s}}(\alpha_k,\lambda_k)=\frac{i}{\sqrt{2}}\begin{pmatrix}
1-3(\tilde{\gamma}_M-2)\left[2\alpha_k^2+2\alpha_k\lambda_k-(\varrho-1)\lambda_k\right]\\[2ex]
1-3(\tilde{\gamma}_M-2)\left[2\lambda_k^2+2\alpha_k\lambda_k-\left(\frac{1}{\varrho}-1\right)\alpha_k\right]

\end{pmatrix}+\mathcal{O}(4),
\ee
where
\be
J_1=\frac{J_{\text{t}}}{Z}, \qquad J_2=1+\mathcal{O}(2).
\ee
The inverse of $\mathcal{J}$ is
\be\label{eq:Jinvmatrix}
\mathcal{J}^{-1}(\alpha,\lambda)=\begin{pmatrix}
\frac{\alpha}{J_1 Z^3}& \frac{\varrho\lambda}{J_1Z^3} \\[2ex]
\frac{-\sqrt{\varrho}Z}{J_2\alpha} \ \ \ \ & \frac{\sqrt{\varrho}Z}{J_2\lambda}
\end{pmatrix}.
\ee
Then the perturbations that are conserved in the superhorizon limit read
\be\label{eq:Rc}
\mathcal{R}^{\text{C}}=\frac{\mathcal{C}_{\text{s}}}{k^{3/2}}\mathcal{J}^{-1}(\alpha,\lambda)\eta w_{\text{s}}, \qquad \mathcal{C}_{\text{s}}=\frac{C_{\text{t}}}{\sqrt{3}}
\begin{pmatrix}
1 \ \  \ \ \ & 0 \\
0 \ \  \ \ \ &1/\varrho
\end{pmatrix},
\ee
where the constant matrix $\mathcal{C}_{\text{s}}$ has been fixed so the first entry of $\mathcal{R}^{\text{C}}$ reproduces $\mathcal{R}_{\chi}$ at $\phi=0$ (and $\mathcal{R}_{\phi}$ at $\chi=0$) and such that $\mathcal{R}^{\text{C}}$ is explicitly invariant under~\eqref{eq:alphalambdaexch}.
Defining the power-spectrum matrix elements $\mathcal{P}_{ij}$ as
\be
\langle \hat{\mathcal{R}}^{\text{C}}_{\mathbf{k}i}\hat{\mathcal{R}}^{\text{C}}_{\mathbf{k}^{\prime
}j}\rangle =(2\pi )^{3}\delta ^{(3)}(\mathbf{k}+\mathbf{k}%
^{\prime })\frac{2\pi ^{2}}{k^{3}}\mathcal{P}_{ij},
\ee
in the superhorizon limit we find
\be
\mathcal{P}_{11}=\frac{C^2_{\text{t}}}{6\pi^2}\left[(\mathcal{J}_{k,11}^{-1})^2|\tilde{Q}_{0\text{s},1}|^2+(\mathcal{J}^{-1}_{k,12})^2|\tilde{Q}_{0\text{s},2}|^2\right],
\ee
\be
\mathcal{P}_{22}=\frac{C^2_{\text{t}}}{6\pi^2\varrho^2}\left[(\mathcal{J}_{k,21}^{-1})^2|\tilde{Q}_{0\text{s},1}|^2+(\mathcal{J}^{-1}_{k,22})^2|\tilde{Q}_{0\text{s},2}|^2\right],
\ee
\be
\mathcal{P}_{12}=\frac{C^2_{\text{t}}}{6\pi^2\varrho}\left[\mathcal{J}_{k,11}^{-1}\mathcal{J}_{k,21}^{-1}|\tilde{Q}_{0\text{s},1}|^2+\mathcal{J}^{-1}_{k,12}\mathcal{J}_{k,22}^{-1}|\tilde{Q}_{0\text{s},2}|^2\right],
\ee
where $\mathcal{J}^{-1}_{k,ij}$ are the matrix elements of $\mathcal{J}^{-1}(\alpha_k,\lambda_k)$. At the leading order the components of $\mathcal{P}_{ij}$ are
\be
\begin{split}
\mathcal{P}_{11}=\frac{m_{\chi}^2\kappa^2}{54(4\pi^2)}\frac{1}{(\alpha_k^2+\varrho\lambda_k^2)^3}&\left[\alpha_k^2+\varrho^2\lambda_k^2+\mathcal{O}(4)\right],
\end{split}
\ee
\be
\begin{split}
\mathcal{P}_{22}=\frac{m_{\chi}^2\kappa^2}{54(4\pi^2)}\frac{(\alpha_k^2+\lambda_k^2)}{\alpha_k^2\lambda_k^2}\left[\alpha_k^2+\varrho\lambda_k^2+\mathcal{O}(4)\right],
   \end{split}
\ee
\be
\begin{split}
\mathcal{P}_{12}=\frac{m_{\chi}^2\kappa^2}{54(4\pi^2)}\frac{\varrho-1}{\sqrt{\varrho}(\alpha_k^2+\varrho\lambda_k^2)}\left[1+\mathcal{O}(2)\right].
   \end{split}
\ee
Since $0<\alpha_k,\lambda_k<1$, the leading entry of $\mathcal{P}_{ij}$ is $\mathcal{P}_{11}$,which we can write up to the NL order
\be
\begin{split}
\mathcal{P}_{11}=&\frac{m_{\chi}^2\kappa^2}{54(4\pi^2)}\frac{1}{(\alpha_k^2+\varrho\lambda_k^2)^3}\big[\alpha_k^2+\varrho^2\lambda_k^2+(17-12\gamma_M)(\alpha_k^4+\varrho^2\lambda_k^4)\\
&+12(2-\gamma_M)(\alpha_k^3\lambda_k+\varrho^2\alpha_k\lambda_k^3)+\alpha_k^2\lambda_k^2(5-24 \varrho +5 \varrho ^2-6 \gamma _M \left(\varrho-1\right)^2)+\mathcal{O}(6)\big],
\end{split}
\ee
 Therefore, we choose $\mathcal{R}^{\text{C}}_1$ and $\mathcal{R}^{\text{C}}_2$ as alternative gauge-invariant quantities instead of the adiabatic and isocurvature perturbations, respectively. 

The spectral indices $n_{ij}(k)\equiv1+\frac{\mathrm{d}\ln\mathcal{P}_{ij}}{\mathrm{d}\ln k}$ are given by
\be
n_{ij}(k)-1=-\beta_{\alpha}(\alpha_k,\lambda_k)\frac{\partial\ln\mathcal{P}_{ij}}{\partial\alpha_k}-\beta_{\lambda}(\alpha_k,\lambda_k)\frac{\partial\ln\mathcal{P}_{ij}}{\partial\lambda_k}
\ee
and to the leading order they read
\be\label{eq:ns}
n_{11}(k)-1= -\frac{6 \left[2\alpha_k ^4+2\lambda_k ^4 \varrho ^2+ \alpha_k ^2 \lambda_k ^2 \left(1+\varrho\right)^2\right]}{\alpha_k ^2+\lambda_k ^2 \varrho
   ^2}+\tilde{\mathcal{O}}(4),
\ee

\be
n_{22}(k)-1=-\frac{6\left(\varrho-1\right)}{\varrho}\frac{\alpha_k^4-\varrho \lambda_k^4}{\alpha_k^2+\lambda_k^2}+\tilde{\mathcal{O}}(4),
\ee

\be
n_{12}(k)-1=-6\left(\alpha_k^2+\lambda_k^2\right)+\tilde{\mathcal{O}}(4),
\ee
where the corrections denoted by $\tilde{\mathcal{O}}(N)$ are of the form
\be
\frac{P_n(\alpha_k,\lambda_k)}{(\alpha_k^2+\varrho^2\lambda_k^2)^m}, \qquad n+2m=N,
\ee
where $P_n$ is a polynomial of degree $n$.

Finally, using the tensor power spectrum~\eqref{eq:Pt} the tensor-to-scalar ratio reads
\be\label{eq:ral}
r(k)\equiv \frac{\mathcal{P}_{\text{t}}(k)}{\mathcal{P}_{11}(k)}=48\frac{(\alpha_k^2+\varrho \lambda_k^2)^2}{\alpha_k^2+\varrho^2 \lambda_k^2}+\tilde{\mathcal{O}}(4),
\ee
where, again, the corrections are not polynomial in $\alpha_k$ and $\lambda_k$.

If we compare~\eqref{eq:ns} and~\eqref{eq:Nk} we get
\be
n_{11}(k)-1=-\frac{2}{N(k)}-\frac{6\alpha_k^2\lambda_k^2(\varrho-1)^2}{\alpha_k^2+\varrho^2\lambda_k^2}.
\ee
Therefore, the double-field dynamics modifies the relation between $N(k)$ and $n_{11}$. On the other hand, comparing~\eqref{eq:ral} with~\eqref{eq:Nk} we see that 
\be
N(k)\simeq\frac{8}{r(k)}
\ee
at the leading order, which coincides with the single-field relation in the case of a quadratic potential. This was already noticed in~\cite{Kim:2006ys} using the $\delta N$ formalism~\cite{Sasaki:1995aw}. However, the conserved gauge-invariant perturbation build in~\autoref{subsec:scalarpert} and used to derive the dominant component of the scalar power-spectrum matrix  does not coincide with that of~\cite{Kim:2006ys}. Using~\eqref{eq:Jinvmatrix} and~\eqref{eq:Rc} we can write $\mathcal{R}_{1}^{\text{C}}$ in terms of the gauge-invariant perturbations~\eqref{eq:RchiRphi} 
\be
\mathcal{R}^{\text{C}}_1=\frac{\dot{\bar{\chi}}^2}{\dot{\bar{\chi}}^2+\varrho\dot{\bar{\phi}}^2}\mathcal{R}_{\chi}+\frac{\varrho\dot{\bar{\phi}}^2}{\dot{\bar{\chi}}^2+\varrho\dot{\bar{\phi}}^2}\mathcal{R}_{\phi}.
\ee
 while the one used in~\cite{Kim:2006ys} is given by~\cite{Sasaki:1995aw}
\be
\mathcal{R}_{\delta N}=\frac{\partial N}{\partial \bar{\chi}_k}\delta\chi+\frac{\partial N}{\partial \bar{\phi}_k}\delta\phi=\frac{\partial N}{\partial \bar{\chi}_k}\frac{\dot{\bar{\chi}}}{H}\mathcal{R}_{\chi}+\frac{\partial N}{\partial \bar{\phi}_k}\frac{\dot{\bar{\phi}}}{H}\mathcal{R}_{\phi},
\ee
where in the last step we have used the spatially-flat gauge. However, at the leading order in the slow-roll approximation the two perturbations coincide. This can be shown in the following way. First, formula~\eqref{eq:Nk} in terms of background fields gives 
\be
N(k)\simeq \frac{\kappa^2}{4}\left(\bar{\chi}_k^2+\bar{\phi}_k^2\right),
\ee
where we have used that at the leading order in $\alpha_k$ and $\lambda_k$
\be
\bar{\chi}_k\simeq-\frac{2}{3 \hat{\kappa}}\frac{\alpha_k}{\alpha_k^2+\varrho\lambda_k^2}, \qquad \bar{\phi}_k\simeq-\frac{2}{3 \hat{\kappa}}\frac{\varrho\lambda_k}{\alpha_k^2+\varrho\lambda_k^2}.
\ee
Then we have
\be
\mathcal{R}_{\delta N}\simeq\frac{\dot{\bar{\chi}}_k\dot{\bar{\chi}}}{\dot{\bar{\chi}}_k^2+\varrho \dot{\bar{\phi}}_k^2}\mathcal{R}_{\chi}+\frac{\varrho\dot{\bar{\phi}}_k\dot{\bar{\phi}}}{\dot{\bar{\chi}}_k^2+\varrho \dot{\bar{\phi}}_k^2}\mathcal{R}_{\phi}.
\ee
Once the background quantities are expanded in powers of $\alpha_k$ and $\lambda_k$, at the leading order we have
\be
\mathcal{R}_{\delta N}\simeq \mathcal{R}^{\text{C}}_1.
\ee 
Nevertheless, the perturbation $\mathcal{R}_{\delta N}$ is not conserved on superhorizon scales. This becomes evident at the NL order in $\alpha_k$  and $\lambda_k$.

Finally, as mentioned in~\autoref{sec:background}, it is convenient to rewrite the observables in terms of the following variables 
\be\label{eq:xy}
x_k\equiv\frac{\alpha_k^2}{\sqrt{\alpha_k^2+\varrho^2\lambda_k^2}}, \qquad y_k\equiv\frac{\varrho\lambda_k^2}{\sqrt{\alpha_k^2+\varrho^2\lambda_k^2}},
\ee
that replace $\alpha_k$ and $\lambda_k$. In fact, using~\eqref{eq:xy} all the relevant quantities can be written as power series, times overall functions
\
\be
\begin{split}
\mathcal{P}_{11}(k)=&\frac{m_{\chi}^2\kappa^2}{54(4\pi^2)(x_k+y_k)^3(x_k+\varrho y_k)}\Bigg[1+\left(17-12\gamma_M\right)(x_k^2+y_k^2)\\
&-x_ky_k\big(24+6\gamma_M(\zeta-2)-5\zeta\big)-\frac{12}{\sqrt{\varrho}}\sqrt{x_ky_k}(x_k+\varrho y_k)(\gamma_M-2)+\mathcal{O}(4)\Bigg],
\end{split}
\ee
\be
n_{11}(k)-1=-12\left[x_k^2+y_k^2+x_ky_k\left(1+\frac{\zeta}{2}\right)\right]+\mathcal{O}(4),
\ee

\be
r(k)=48(x_k+y_k)^2\left\{1+6(\gamma_M-2)\left[(x_k-y_k)^2+\frac{2\sqrt{x_ky_k}}{\sqrt{\varrho}}(x_k+\varrho y_k)\right]\right\}+\mathcal{O}(6),
\ee

\be
\beta_{\mathcal{P}}(k)=\frac{(x_k+y_k)^3(x_k^2+y_k^2+\zeta x_ky_k)}{x_ky_k}\left[x_k+y_k+\mathcal{O}(3)\right],
\ee
where the quantity $\beta_{\mathcal{P}}$ is defined as the ratio
\be
\beta_{\mathcal{P}}\equiv\frac{\mathcal{P}_{22}}{\mathcal{P}_{11}+\mathcal{P}_{22}},
\ee
analogous to $\beta_{\text{iso}}$ used in~\cite{Akrami:2018odb}.

For completeness we write the tensor power spectrum and its spectral index using the variables~\eqref{eq:xy}.  To the leading order we have
\be
\mathcal{P}_{\text{t}}\simeq\frac{2m_{\chi}^2\kappa^2}{9\pi^2}\frac{1}{(x_k+y_k)(x_k+\varrho y_k)},\qquad n_{\text{t}}\simeq-6(x_k^2+y_k^2+\zeta x_ky_k).
\ee
It is worth to note that the single-field relation $r+8n_{\text{t}}\simeq 0$ is violated already at the leading order. Indeed
\be
r+8n_{\text{t}}\simeq -48 x_ky_k(\zeta-2).
\ee
To summarize, we can extract meaningful predictions from a double-field model by means of the perturbative technique presented here. In particular, it is possible to single out a dominant scalar perturbation that is conserved in the superhorizon limit, which makes its power spectrum a good candidate to be compared with experimental data. In the case of the present model the relation between the number of e-folds and the tensor-to-scalar ratio reduces to that of the single-field case (at the leading order).
%%%%%%%%%%%%%%%%%%%%%%%%%%%%%%%%%%%%%%%%%%%%%%%%%%%%%%%%%%%%%%%%%%%%%%%%%%%%%%%%%%%%%%
\sect{Conclusions}
\label{sec:conclusions}
We have developed a perturbative method that allows to discuss multicomponent inflation models in a consistent manner. The strategy enables a systematic study of inflationary dynamics order by order in perturbation expansion in powers of small quantities, which are analogues of small slow-roll parameters in the conventional description.
The method is illustrated by a simple model of two-field inflation possessing only mass terms as the potential. The model properly and nicely reveals various theoretical aspects of the perturbative strategy, even though it is not phenomenologically viable. We have computed the tensor and scalar power spectra to the NNL and NL orders, respectively, although it is straightforward to obtain results up to an arbitrary order. In the case considered here contributions to the inflation dynamics that originate from each inflaton are of the same order, so that this is indeed a truly two-component inflation. It is the case for the background dynamics and it remains true for perturbations as well.
In particular, thanks to nonperturbative resummations, multi-field modifications are present already at the leading order, both in the tensor and scalar power spectra.

We have derived new independent gauge-invariant scalar perturbations that are conserved on superhorizon scales. 
As it is very well known, the single-field scenario with quadratic potential is excluded by the present experimental data. It turns out that, at the leading order, in the presence of a second field with quadratic potential, the relation between the number of e-folds $N(k)$ and the tensor-to-scalar ratio $r(k)$ is the same as for the single-field quadratic inflation. On the other hand the relation between $N(k)$ and the spectral index $n_{11}$ is modified. Nevertheless, the modification of the single-field quadratic inflation are not sufficient to withstand comparison with the Planck~\cite{Planck:2018jri} and  BICEP/Keck~\cite{BICEP:2021xfz} collaborations.

%%%%%%%%%%%%%%%%%%%%%%%%%%%%%%%%%%%%%%%%%%%%%%%%%%%%%%%%%%%%%%%%%%%%%%%%%%%%%%%%%%%%%%
\section*{Acknowledgements}
Authors thank Anish Ghoshal for his interest at the early stage of this project. M.P. thanks A. Melis for useful discussions.
The work of B.G. was supported in part by the National Science Centre (Poland) within the research
project 2023/49/B/ST2/00856.  

\appendix
\renewcommand{\thesection}{\Alph{section}} \renewcommand{\theequation}{%
\thesection.\arabic{equation}} \setcounter{section}{0}

\section{Validity of asymptotic series}
\label{app:asymp}

In this appendix we comment on the asymptotic series used throughout the paper. For concreteness, our discussion is restricted to the expansions of $\chi$, $\phi$ and $H$, but what explained below applies to any expansion derived in the previous sections. 

An asymptotic series is not convergent since its coefficients grow. However, it can still be used to give a good approximation of a function provided that the expansion parameter is small enough and the series is optimally truncated. For simplicity, we consider an asymptotic series $S(x)$ around $x=0$
\be
S(x)=\sum_{n=0}^{\infty}a_nx^n
\ee
and define the quantity
\be\label{eq:bn}
b_n(x)\equiv a_nx^n.
\ee
Then, given a value $x_0$ for $x$, a good criterion to determine the order of the optimal truncation is to find the maximum $n$ for which the ratio
\be\label{eq:ratiob}
\left|\frac{b_{n+1}(x_0)}{b_n(x_0)}\right|<1.
\ee
We label such order as $N_{\text{opt}}(x_0)$.
\begin{figure}
    \centering
    \includegraphics[width=1
\linewidth]{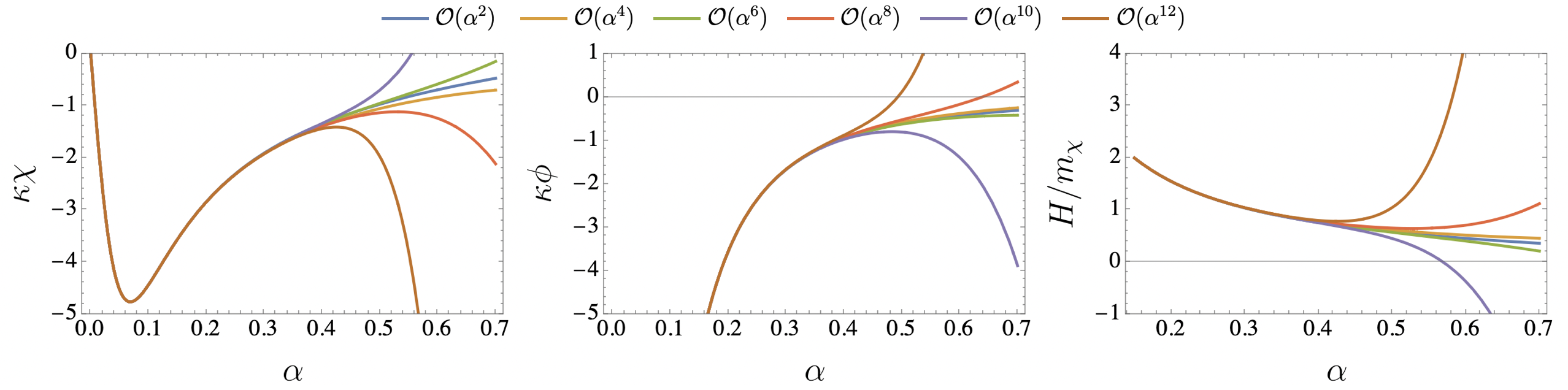}
    \caption{The three plots show the expansions of $\kappa\chi$, $\kappa\phi$ and $H/m_{\chi}$ as functions of $\alpha$ for  $\lambda=0.02$ and $\varrho=12$. The expansions are truncated at various orders to show their behavior as asymptotic series. The order of the truncation is shown in the legends.}
    \label{fig:truncations}
\end{figure}

In our case every series is an expansion in two variables $\alpha$ and $\lambda$. In order to show the asymptotic nature of the series, we first truncate them at various orders $\mathcal{O}(N)$, then fix $\lambda$ and $\varrho$ to some values. The results are shown in~\autoref{fig:truncations} where  $\kappa\chi$, $\kappa\phi$ and $H/m_{\chi}$ are plotted as functions of $\alpha$ for fixed $\lambda=0.02$ and $\varrho=12$. One can see that for small values of $\alpha$ all truncations predict similar results, while as $\alpha$ increases they increasingly depart form each other. A typical behavior of asymptotic series is that higher-order truncations deviate for lower values of $\alpha$ and by a larger amount compared to the deviations between lower-order ones. This is particularly evident in the second panel, where the first 3 truncations remain comparable up to $\alpha\sim 0.6$, while the others depart well before.

Now we check the criterion for the optimal truncation explained above by choosing different values of $\alpha$. Since the expansions~\eqref{eq:chitildeseries},  \eqref{eq:phitildeseries} and~\eqref{eq:Htildeseries} 
have only even powers of $\alpha$, we set $x=\alpha^2$ in~\eqref{eq:bn}.

For $\lambda=0.02$ and $\varrho=12$ we have 
\be
\chi:\qquad N_{\text{opt}}(0.2)>12, \qquad N_{\text{opt}}(0.3)=10, \qquad N_{\text{opt}}(0.4)=6,\qquad N_{\text{opt}}(0.5)=4,
\ee
\be
\phi:\qquad N_{\text{opt}}(0.2)>12, \qquad N_{\text{opt}}(0.3)=8, \qquad N_{\text{opt}}(0.4)=6,\qquad N_{\text{opt}}(0.5)=2,
\ee
\be
H:\qquad N_{\text{opt}}(0.2)>12, \qquad N_{\text{opt}}(0.3)=10, \qquad N_{\text{opt}}(0.4)=6,\qquad N_{\text{opt}}(0.5)=4.
\ee
Therefore, in this case, for $\alpha<0.2$ it is safe to truncate the series at order 12 or even higher. However, it is worth to highlight that the higher-order corrections will eventually start to compete with loop corrections.
%%%%%%%%%%%%%%%%%%%%%%%%%%%%%%%%%%%%%%%%%%%%%%%%%%%%%%%%%%%%%%%%%%%%%%%%%%%%%%%%%%%%%%

\bibliographystyle{JHEP}
\bibliography{ref}

\end{document}